\documentclass[12pt,thmsa]{article}
\usepackage{amssymb}

\usepackage{sw20lart}

\input{tcilatex}
\begin{document}

\begin{center}
{\Large Efficient multiple-quantum transition processes in an n-qubit spin
system }

\bigskip {\large Author: Xijia Miao \\[0pt]
Date: June, 2004}\\[0pt]
Somerville, Massachusetts

\vspace{0.15in}\vspace{0.1in}{\large Abstract}
\end{center}

The whole Hilbert state space of an $n-$qubit spin system can be divided
into $n+1$ state subspaces according to the angular momentum theory of
quantum mechanics. Here it is shown that any unknown state in such a state
subspace, whose dimensional size is proportional to either a polynomial or
exponential function of the qubit number $n$, can be transferred efficiently
into a larger subspace with a dimensional size generally proportional to an
exponential function of the qubit number by the multiple-quantum unitary
transformation with a subspace-selective multiple-quantum unitary operator.
The efficient quantum circuits for the subspace-selective multiple-quantum
unitary operators are really constructed. \newline
\newline
{\large 1. Multiple-quantum transition between state subspaces of the
Hilbert space}

Multiple-quantum transition processes [1] are closely related to the quantum
computation [2, 3]. In a quantum system such as a coupled multiple-particle
two-level quantum system, e.g., an $n-$qubit spin system which may consist
of $n$ coupled nuclei of spin $1/2,$ quantum state transfer or transition
between different states of the Hilbert space of the system generally
consists of a variety of multiple-quantum transition processes except those
noncoherent transition processes which are usually non-unitary processes
[1]. A quantum search algorithm [4] running in a quantum system usually
starts at an initial state such as a superposition, then performs a sequence
of known unitary operations and the oracle unitary operation to convert the
initial state into the marked state in a high efficiency, and finally makes
a quantum-mechanically measure to output the marked state. A variety of
quantum state transition or transfer from one state to another really occur
in the quantum system during the quantum search process. The quantum search
problem usually takes the whole Hilbert space of a quantum system as the
search state space in which the unknown marked state is searched for. The
computational complexity of the quantum search problem is closely related to
the complexity of the multiple-quantum transition processes between
different subspaces of the Hilbert space of the $n-$qubit spin system. The
quantum state transition or transfer generally could not be efficient for an
unknown state, e.g., the marked state in the quantum search problem from a
state subspace of the Hilbert space whose dimensional size is proportional
to an exponential function of qubit number $n$ of the system into a smaller
state subspace whose dimensional size increases polynomially as the qubit
number, because a quantum state in a known subspace with a polynomial
dimensional size can be determined in polynomial time. The multiple-quantum
unitary transformation has been used to help design of quantum circuit [5]
and quantum algorithms in quantum computation [2, 3]. The multiple-quantum
spectra could be used to output results of quantum computation [3]. Using
multiple-quantum unitary transformation the quantum search space of the
quantum search problem [4] could be reduced from the whole Hilbert space of
the $n-$qubit spin system to its largest subspace whose dimensional size is
still much smaller than that of the whole Hilbert space, and hence this
reduction could speed up the quantum search process. The more important is
that the multiple-quantum unitary transformation provides a useful method to
manipulate quantum state transition or transfer from a state to another or
from a subspace to another in the Hilbert space in quantum computation.

The Hilbert space of an $n-$qubit spin system has a dimensional size $2^{n}$
which increases exponentially as the qubit number $n$. It has $2^{n}$
conventional computational bases that can be used to represent $2^{n}$
numbers or elements in quantum computation. This is a large search space in
the quantum search problem. The quantum search space will be reduced if the
whole Hilbert space can be divided into small subspaces. In quantum
computation this subspace reduction for the Hilbert space has been proposed
to help design for the quantum search algorithm [2, 6]. There are a number
of symmetric property of a quantum system to achieve the decomposition for
its Hilbert space. The important one is the rotation symmetry in space of a
quantum system [7-10]. The rotation symmetry in spin space of an $n-$qubit
spin system which consists of $n$ spins$-1/2$ may be used to guide the
decomposition for the Hilbert space into its small subspaces. The angular
momentum theory of quantum mechanics gives a detailed description for the
rotation symmetry in space of a quantum system [7-10], and according to the
angular momentum theory the whole Hilbert space of the $n-$qubit spin system
may be divided into $(n+1)$ state subspaces, and each of which may be formed
from the complete set of the eigenstates $\{|\Psi _{M}\rangle \}$ of the
total magnetic quantum operator $I_{z}=\sum_{k=1}^{n}I_{kz}$ with a common
eigenvalue $M$ which satisfy the eigen-equation: 
\begin{equation}
I_{z}|\Psi _{M}\rangle =M|\Psi _{M}\rangle ,\text{ }(\hslash =1).  \label{1}
\end{equation}
The total magnetic quantum number (or the eigenvalue) $M$ has $(n+1)$
different values and can take $n/2,n/2-1,......,-n/2,-n/2$ for the $n-$qubit
spin system$.$ Each value of the quantum number $M$ remarks a state
subspace. The state subspace with the total magnetic quantum number $M=n/2-k$
is denoted as $S(M=n/2-k)$ or simply as $S_{zq}(k)$ with $k=0,1,...,n.$
Since all the states in a subspace take the same value of the total magnetic
quantum number there is not a change for the value of the total magnetic
quantum number in quantum transition between different states within the
subspace and hence the quantum transition is a zero--quantum transition. A
quantum state transition from a subspace to another is a nonzero-order
quantum transition since the total magnetic quantum number value is changed
when a state is transferred from a subspace into another. According to the
angular momentum theory the dimensional size for the subspace $S_{zq}(k)$
with $k=0,1,...,n$ is given by $\left( 
\begin{array}{l}
n \\ 
k
\end{array}
\right) $ for the $n-$qubit spin system$,$ which is denoted as $d(M=n/2-k)$
or simply as $d(k).$ Among the $(n+1)$ subspaces the two smallest subspaces
are $S_{zq}(0)$ and $S_{zq}(n)$, whose dimension is one. The next smallest
subspaces are $S_{zq}(1)$ and $S_{zq}(n-1)$ which have the same dimensional
size $n$. For a spin system with an even qubit number $n$ the largest
subspace is $S_{zq}(n/2)$ and its dimensional size is $%
d(n/2)=n!/[(n/2)!]^{2} $. For a spin system with an odd qubit number the two
largest subspaces are $S_{zq}((n-1)/2)$ and $S_{zq}((n+1)/2),$ respectively$%
. $ Both the two subspaces have the same dimensional size equal $%
d((n-1)/2)=n!/\{[(n-1)/2]![(n+1)/2]!\}.$ For a large $n$ the number $%
d(n/2)\thickapprox d((n-1)/2)\thickapprox 2^{n}/\sqrt{\pi n/2}$ by the
Starling$^{\prime }s$ formula. Therefore, the dimensional size for the
largest subspace increases exponentially as the qubit number $n$. Since any
computational base of the Hilbert space of the spin system can be only in
one of these $(n+1)$ subspaces the quantum search space now may be limited
to such a subspace in which the marked state is. When the marked state is in
those smallest subspaces such as $S_{zq}(0)$, $S_{zq}(1)$, $S_{zq}(2)$,
etc., whose dimensional size increases polynomially as the qubit number, it
may be found in polynomial time. However, if the marked state is in the
largest subspaces then it can be found by the Grover quantum search
algorithm [4] but this need take an exponential time proportional to $\sqrt{%
d(n/2)}$. Therefore, this shows indirectly that an unknown state usually
could not be transferred efficiently from one large subspace whose
dimensional size increases exponentially as the qubit number into a small
subspace with a dimensional size proportional to a polynomial function of
the qubit number. Actually, this quantum-state transfer is closely related
to the computational complexity of the quantum search problem [4], that is,
the quantum-state transfer is as hard as the latter one. However, the
inverse quantum-state transfer process could be efficient, that is, an
unknown state in a small subspace whose dimensional size may be proportional
to an exponential or polynomial function of the qubit number could be
efficiently transferred into a larger subspace by a multiple-quantum unitary
transformation. These quantum-state transfer processes involve in
quantum-state multiple-quantum transitions between different subspaces. It
seems to see clearly that the quantum-state transfer could be efficient if
an unknown state is initially in those smallest subspaces such as $S_{zq}(1)$%
, $S_{zq}(2)$, etc. However, it is not yet clear whether the quantum-state
transfer is efficient or not if an unknown state is initially in those small
subspaces whose dimensional size is also proportional to an exponential
function of the qubit number. In the following a subspace-selective
multiple-quantum unitary operator is constructed that transfers completely
an unknown state from a subspace into a larger subspace of the Hilbert space.

Suppose that an unknown state $|\Psi _{s}\rangle $ is in a state subspace $%
S(M_{s})$ and needs to be transferred to a larger subspace $S(M_{s}+p)$ by a
multiple-quantum unitary transformation. The state $|\Psi _{s}\rangle $ can
be expanded conveniently in terms of the usual computational basis $%
\{|\varphi _{k}(M_{s})\rangle \}$ of the subspace $S(M_{s}):$%
\begin{equation}
|\Psi _{s}\rangle =\sum_{k=0}^{d(M_{s})-1}a_{sk}|\varphi _{k}(M_{s})\rangle .
\label{2}
\end{equation}
If there is a $p-$quantum unitary operator that can convert simultaneously
all the computational bases $\{|\varphi _{k}(M_{s})\rangle \}$ in the
unknown state $|\Psi _{s}\rangle $ from the subspace $S(M_{s})$ into the
subspace $S(M_{s}+p)$ then this $p-$quantum unitary operator also can
convert the unknown state $|\Psi _{s}\rangle $ (2) from the subspace $%
S(M_{s})$ into the subspace $S(M_{s}+p)$. It is possible to construct such a 
$p-$quantum unitary operator. Now dimensional sizes for the subspaces $%
S(M_{s})$ and $S(M_{s}+p)$ are $d(M_{s})$ and $d(M_{s}+p),$ respectively,
and hence there are $d(M_{s})$ and $d(M_{s}+p)$ computational bases in the
subspaces $S(M_{s})$ and $S(M_{s}+p)$, respectively. By using the
computational bases of the two subspaces one can generate a
subspace-selective $p-$quantum unitary operator that converts the unknown
state $|\Psi _{s}\rangle $ from the subspace $S(M_{s})$ into $S(M_{s}+p).$
Because $d(M_{s}+p)$ is greater than $d(M_{s})$ one can choose properly any $%
d(M_{s})$ bases among the $d(M_{s}+p)$ bases of the subspace $S(M_{s}+p)$.
By combining these $d(M_{s})$ computational bases from the subspace $%
S(M_{s}+p)$ with all the $d(M_{s})$ computational bases of the subspace $%
S(M_{s})$ one can build up $d(M_{s})$ state-selective $p-$quantum Hermitian
operators. For every pair of the computational bases $|\varphi
_{k}(M_{s})\rangle $ of the subspace $S(M_{s})$ and $|\varphi
_{k}(M_{s}+p)\rangle $ of the subspace $S(M_{s}+p)$ a state-selective $p-$%
quantum Hermitian operator is built up: 
\begin{equation}
Q_{psk}=\frac{1}{2}(|\varphi _{k}(M_{s})\rangle \langle \varphi
_{k}(M_{s}+p)|+|\varphi _{k}(M_{s}+p)\rangle \langle \varphi _{k}(M_{s})|).
\label{3}
\end{equation}
Then the corresponding state-selective $p-$quantum unitary operator is
generated from the Hermitian operator $Q_{psk}$ by 
\begin{equation}
U_{psk}(\theta )=\exp (-i\theta Q_{psk}).  \label{4}
\end{equation}
This state-selective $p-$quantum unitary operator is only applied to both
the two states $|\varphi _{k}(M_{s})\rangle $ of the subspace $S(M_{s})$ and 
$|\varphi _{k}(M_{s}+p)\rangle $ of the subspace $S(M_{s}+p).$ There are $%
d(M_{s})$ such state-selective $p-$quantum unitary operators. All these
state-selective $p-$quantum unitary operators are commutable with each other
because all the computational bases are orthogonal to each other. Then by
multiplying all these $d(M_{s})$ state-selective $p-$quantum unitary
operators a subspace-selective $p-$order quantum unitary operator is
obtained: 
\begin{equation}
U_{ps}(\theta )=\prod_{k=0}^{d(M_{s})-1}U_{psk}(\theta )=\exp (-i\theta 
\stackrel{d(M_{s})-1}{\stackunder{k=0}{\sum }}Q_{psk}).  \label{5}
\end{equation}
This $p-$quantum unitary operator is selectively applied to both the two
subspaces $S(M_{s})$ and $S(M_{s}+p).$ Since the dimensional size $d(M_{s})$
of the subspace $S(M_{s})$ may increase exponentially as the qubit number
the unitary operator $U_{ps}(\theta )$ (5) may contain exponentially many
state-selective $p-$quantum unitary operators (4). First it can be proved
that the basis state $|\varphi _{k}(M_{s})\rangle $ of the subspace $%
S(M_{s}) $ can be converted completely into the basis state $|\varphi
_{k}(M_{s}+p)\rangle $ of the subspace $S(M_{s}+p)$ by the state-selective $%
p-$quantum unitary operator $U_{psk}(\theta ).$ By expanding the unitary
operator $U_{psk}(\theta )$ (4)$:$ 
\begin{eqnarray}
U_{psk}(\theta ) &=&E+(-1+\cos \frac{1}{2}\theta )(|\varphi
_{k}(M_{s})\rangle \langle \varphi _{k}(M_{s})|  \nonumber \\
&&+|\varphi _{k}(M_{s}+p)\rangle \langle \varphi
_{k}(M_{s}+p)|)-i2Q_{psk}\sin \frac{1}{2}\theta  \label{6}
\end{eqnarray}
and using the orthogonormal condition for the usual computational bases one
easily obtains \newline
\begin{equation}
U_{psk}(\pi )|\varphi _{s^{\prime }}(M_{s})\rangle =\left\{ 
\begin{array}{l}
-i|\varphi _{s^{\prime }}(M_{s}+p)\rangle ,\text{ if }s^{\prime }=k \\ 
|\varphi _{s^{\prime }}(M_{s})\rangle ,\text{ if }s^{\prime }\neq k
\end{array}
\right. .  \label{7}
\end{equation}
Therefore, the basis state $|\varphi _{k}(M_{s})\rangle $ is transferred
completely to the basis state $|\varphi _{k}(M_{s}+p)\rangle $ in additional
to a phase factor $-i$ by the $p-$quantum unitary operator $U_{psk}(\pi )$.
However, any other basis state of the subspace $S(M_{s})$ keeps unchanged
under the $p-$quantum unitary operator. This result can be further used to
prove that any computational basis state $|\varphi _{s^{\prime
}}(M_{s})\rangle $ of the subspace $S(M_{s})$ can be converted completely
into the basis state $|\varphi _{s^{\prime }}(M_{s}+p)\rangle $ of the
subspace $S(M_{s}+p)$ by the subspace-selective $p-$quantum unitary operator 
$U_{ps}(\pi )$ in (5)$.$ Because all the state-selective $p-$quantum unitary
operators $Q_{psk}$ $(k=0,1,...,d(M_{s})-1)$ are commutative to each other
one therefore has for any basis state $|\varphi _{s^{\prime }}(M_{s})\rangle 
$ of the subspace $S(M_{s})$, by using (7), 
\begin{eqnarray}
U_{ps}(\pi )|\varphi _{s^{\prime }}(M_{s})\rangle &=&\stackrel{d(M_{s})-1}{%
\stackunder{k=0}{\prod }}\exp (-i\pi Q_{psk})|\varphi _{s^{\prime
}}(M_{s})\rangle  \nonumber \\
&=&\exp (-i\pi Q_{pss^{\prime }})|\varphi _{s^{\prime }}(M_{s})\rangle 
\nonumber \\
&=&-i|\varphi _{s^{\prime }}(M_{s}+p)\rangle .  \label{8}
\end{eqnarray}
Thus, the basis state $|\varphi _{s^{\prime }}(M_{s})\rangle $ is completely
transferred into the basis state $|\varphi _{s^{\prime }}(M_{s}+p)\rangle $
of the subspace $S(M_{s}+p)$ in addition to a total phase factor $-i$ by the 
$p-$quantum unitary operator $U_{ps}(\pi )$. This also indicates that all
the basis states of the subspace $S(M_{s})$ in the unknown state $|\Psi
_{s}\rangle $ (2) can be simultaneously converted completely into the
subspace $S(M_{s}+p)$ by the $p-$quantum unitary operator $U_{ps}(\pi ).$
Then an arbitrary state of the subspace $S(M_{s})$ also including $|\Psi
_{s}\rangle $ with the expansion (2) can be also transferred completely into
the subspace $S(M_{s}+p)$ by the $p-$quantum unitary operator $U_{ps}(\pi ).%
\newline
$

Generally, it is hard to construct an efficient subspace-selective
multiple-quantum unitary operator that applies selectively on both two
subspaces with exponentially many basis states. But since the larger
subspace $S(M_{s}+p)$ has a larger dimensional size than the subspace $%
S(M_{s})$ then it could be possible to construct such an efficient
subspace-selective multiple-quantum unitary operator as $U_{ps}(\theta )$
(5) by suitably choosing the $d(M_{s})$ bases $\{|\varphi
_{k}(M_{s}+p)\rangle \}$ among all the $d(M_{s}+p)$ $(\geq d(M_{s}))$
computational bases of the subspace $S(M_{s}+p)$. An important
subspace-selective $p-$quantum unitary operator of (5) is related to both a
small subspace $S_{zq}(k)$ $(k\neq n/2)$ and the largest subspace $%
S_{zq}(n/2)$ of the Hilbert space of an $n-$qubit spin system. If the marked
state in the quantum search problem belongs to a state subspace $S_{zq}(k)$ $%
(k\neq n/2)$ then using the known $p-$quantum unitary operator $U_{ps}(\pi )$
one may transfer efficiently it from the subspace $S_{zq}(k)$ into the
largest subspace $S_{zq}(n/2)$. This really reduces the whole Hilbert space
to its largest subspace $S_{zq}(n/2)$ as the search space of the quantum
search problem. Below it is shown how an efficient quantum circuit for the
subspace-selective $p-$quantum unitary operator (5) can be constructed in an 
$n-$qubit spin system. \newline
\newline
{\large 2.The basic unitary operations}\newline

The basic unitary operations $U(\theta )$ used to build up the
subspace-selective multiple-quantum unitary operators (5) may be generated
from the Hermitian product operators $Q$ by the exponential mapping: $%
U(\theta )=\exp (-i\theta Q).$ The Hermitian product operators are the
tension product operators of single-spin operators in an $n-$qubit spin
system and generally written as $Q=H_{1}\bigotimes H_{2}\bigotimes
...\bigotimes H_{n}$. The Hermitian operator $H_{k}$ is the single-spin
operator of the $k$th spin of the spin system and can be generally expressed
as a linear combination of the single-spin magnetization operators $I_{kx},$ 
$I_{ky},$ and $I_{kz}$ as well as the unity operator $E_{k}$ $:$ $%
H_{k}=\alpha _{k0}E_{k}+\alpha _{kx}I_{kx}+\alpha _{ky}I_{ky}+\alpha
_{kz}I_{kz}$, while the single-spin magnetization operators are related to
Pauli$^{\prime }s$ spin operators $\sigma _{k\mu }$ by $I_{k\mu }=\frac{1}{2}%
\sigma _{k\mu }$ $(\mu =x,y,z).$ Every single-spin operator $H_{k}$ can be
diagonalized: $\tilde{H}_{k}=U_{k}H_{k}U_{k}^{+},$ and its diagonal operator
is given generally by $\tilde{H}_{k}=\alpha _{k}(\frac{1}{2}%
E_{k}+I_{kz})+\beta _{k}(\frac{1}{2}E_{k}-I_{kz}).$ Denote $\tilde{Q}$ as
the diagonal operator of the product operator $Q.$ Then the diagonal
operator is generally written as $\tilde{Q}=\tilde{H}_{1}\bigotimes \tilde{H}%
_{2}\bigotimes ...\bigotimes \tilde{H}_{n}.$ The complete operator set for
the diagonal operators $\tilde{Q}$ forms the $LOMSO$ operator subspace [11].
The complete basis operator set for the $LOMSO$ operator subspace usually
may be chosen conveniently as the longitudinal magnetization and spin order
operator set [11]:

$\tilde{Q}_{A}=\{E,$ $I_{kz},$ $2I_{kz}I_{lz},$ $4I_{kz}I_{lz}I_{mz},$ $...,$
$2^{n-1}I_{1z}I_{2z}...I_{nz}\}.$\newline
In addition to the longitudinal magnetization and spin order operator set
there are also other equivalent complete basis operator sets, of which a
particularly important basis operator set is given by [2]

$\tilde{Q}_{B}=%
\{E,D_{l}^{k_{1}},D_{l}^{k_{1}k_{2}},...,D_{l}^{k_{1}k_{2}...k_{n}}\},$%
\newline
where the diagonal operator $D_{l}^{k_{1}k_{2}...k_{m}}$ $%
(m=1,2,...,n;l=0,1,...,2^{m}-1)$ is defined by 
\begin{eqnarray}
D_{l}^{k_{1}k_{2}...k_{m}} &=&(\frac{1}{2}E_{k_{1}}+a_{k_{1}}^{l}I_{k_{1}z})%
\bigotimes (\frac{1}{2}E_{k_{2}}+a_{k_{2}}^{l}I_{k_{2}z})  \nonumber \\
&&\bigotimes ...\bigotimes (\frac{1}{2}E_{k_{m}}+a_{k_{m}}^{l}I_{k_{m}z}),%
\newline
\label{9}
\end{eqnarray}
where the indices $k_{1},k_{2},...,k_{n}$ and $k,l,m,...,$ are series number
of spins in the spin system and usually are ordered: $1\leq
k_{1}<k_{2}<...<k_{n}\leq n$, $1\leq k<l<...\leq n$; $\{a_{k}^{l}\}$ is a
quantum-state unit-number vector, $a_{k}^{l}=\pm 1;$ and all the unity
operator components $\{E_{k}\}$ in the product operators are omitted for
convenience, for example, the full expressions for the product operators $%
2I_{kz}I_{lz}$ and $D_{l}^{k_{1}k_{2}}$ should be given respectively by

$2I_{kz}I_{lz}\equiv 2E_{1}\bigotimes ...\bigotimes E_{k-1}\bigotimes
I_{kz}\bigotimes E_{k+1}\bigotimes ...$

$\bigotimes E_{l-1}\bigotimes I_{lz}\bigotimes E_{l+1}\bigotimes
...\bigotimes E_{n},$

$D_{l}^{k_{1}k_{2}}\equiv E_{1}\bigotimes ...\bigotimes
E_{k_{1}-1}\bigotimes (\frac{1}{2}E_{k_{1}}+a_{k_{1}}^{l}I_{k_{1}z})%
\bigotimes E_{k_{1}+1}\bigotimes ...$

$\bigotimes E_{k_{2}-1}\bigotimes (\frac{1}{2}%
E_{k_{2}}+a_{k_{2}}^{l}I_{k_{2}z})\bigotimes E_{k_{2}+1}\bigotimes
...\bigotimes E_{n}.$\newline
In particular, denote that the diagonal operator $D_{m,l}\equiv
D_{l}^{12...m}=\stackrel{m}{\stackunder{k=1}{\bigotimes }}(\frac{1}{2}%
E_{k}+a_{k}^{l}I_{kz})$ $(m=1,2,...,n;l=0,1,...,2^{m}-1),$ and there is a
simple denotation $D_{l}\equiv D_{n,l}$ used also in previous papers [2, 6].
The diagonal operator $D_{m,l}$ is called the quantum-state diagonal
operator since a conventional computational basis state of an $m-$qubit spin
system can be characterized completely by the quantum-state unit-number
vector $\{a_{k}^{l}\}$ or equivalently by the diagonal operator $D_{m,l}$
[2].

It is clear that both the basis operator sets $\widetilde{Q}_{A}$ and $%
\widetilde{Q}_{B}$ are equivalent to each other. By expanding the diagonal
operator $D_{l}^{k_{1}k_{2}...k_{m}}$ (9) it can be seen that the operator
is a linear combination of the product operators of the set $\widetilde{Q}%
_{A}.$ On the other hand, using the operator identity:\ $2I_{kz}=(\frac{1}{2}%
E_{k}+I_{kz})-(\frac{1}{2}E_{k}-I_{kz})$ every product operator in the set $%
\widetilde{Q}_{A}$ also can be expressed as a linear combination of the
basis operators of the set $\widetilde{Q}_{B}.$

It is known that elementary propagators built up with the basis operators of
the operator set $\widetilde{Q}_{A}$ can be implemented efficiently. For the
elementary propagators there is a simple recursive relation for decomposing
a multi-body elementary propagator built up with a multi-body interaction
basis operator, e.g., $2^{m-1}I_{k_{1}z}I_{k_{2}z}...I_{k_{m}z},$ into a
sequence of one-, and two-body elementary propagators. Generally, the
elementary propagator $R_{k_{1}k_{2}...k_{m}}(\theta )=\exp (-i\theta
2^{m-1}I_{k_{1}z}I_{k_{2}z}...I_{k_{m}z})$ $(2<m\leq n)$ can be simply
decomposed as [5] 
\begin{eqnarray}
&&R_{k_{1}k_{2}...k_{m}}(\theta )=\exp (-i\frac{\pi }{2}I_{k_{m-1}x})\exp
(-i\pi I_{k_{m-1}z}I_{k_{m}z})  \nonumber \\
&&\times \exp (-i\frac{\pi }{2}I_{k_{m-1}y})\exp (-i\theta
2^{m-2}I_{k_{1}z}I_{k_{2}z}...I_{k_{m-1}z})  \nonumber \\
&&\times \exp (i\frac{\pi }{2}I_{k_{m-1}y})\exp (i\pi
I_{k_{m-1}z}I_{k_{m}z})\exp (i\frac{\pi }{2}I_{k_{m-1}x}).  \label{10}
\end{eqnarray}
This recursive relation assures that any elementary propagator built up with
the basis operator of the operator set $\widetilde{Q}_{A}$ can be
efficiently decomposed into a sequence of one- and two-qubit quantum gates.

The basic unitary operations built up with the basis operators of the
operator set $\tilde{Q}_{B}$ also can be implemented efficiently. It is
based on the fact that the selective rotation operation applying only to a
given state of any $m-$qubit subsystem $(1\leq m\leq n)$ of an $n-$qubit
spin system can be implemented efficiently. It is known that the selective
rotation operation $C_{0}(\theta )=\exp (-i\theta D_{0})$ built up with the
basis operator $D_{0}=\stackrel{n}{\stackunder{k=1}{\bigotimes }}(\frac{1}{2}%
E_{k}+I_{kz})$ of the operator set $\tilde{Q}_{B}$\ can be performed
efficiently in an $n-$qubit spin system [2, 4]. This is an $n-$qubit
selective rotation operation applying only to the known state $|0_{1}\rangle
|0_{2}\rangle ...|0_{n}\rangle $ of the $n-$qubit spin system. Generally, an 
$m-$qubit selective rotation operation with $1\leq m\leq n$ can be also
performed efficiently in the $n-$qubit spin system. This $m-$qubit selective
rotation operation is only applied to the known state $|0_{1}\rangle
|0_{2}\rangle ...|0_{m}\rangle $ of the $m-$qubit subsystem consisting of
the first $m$ spins of the $n-$qubit spin system when it is applied to any
basis state $|0_{1}\rangle |0_{2}\rangle ...|0_{m}\rangle |\varphi
_{m+1}\rangle ...|\varphi _{n}\rangle $ of the $n-$qubit spin system ($%
|\varphi _{m+1}\rangle ,|\varphi _{m+2}\rangle ,...,|\varphi _{n}\rangle $
take $|0\rangle $ or $|1\rangle )$. It is defined by [2] 
\begin{equation}
C_{m,0}(\theta )=\exp (-i\theta D_{m,0}).  \label{11}
\end{equation}
There is a simple denotation: $C_{l}(\theta )=C_{n,l}(\theta )$ also used in
previous papers [2, 6].

The efficient implementation for the $m-$qubit selective rotation operation $%
C_{m,0}(\theta )$ may use conveniently the reversible AND operations and
conditional phase shift operations. The classical AND operation is really
irreversible and it needs to be changed to the reversible one by the Bennett$%
^{\prime }s$ method [12] when it is used to construct these selective
rotation operations.

More generally, any $m-$qubit selective rotation operation $%
C_{l}^{k_{1}k_{2}...k_{m}}(\theta )$ of any $m-$qubit subsystem $(1\leq
m\leq n)$ of the $n-$qubit spin system also can be performed efficiently, 
\begin{equation}
C_{l}^{k_{1}k_{2}...k_{m}}(\theta )=\exp (-i\theta
D_{l}^{k_{1}k_{2}...k_{m}}),  \label{12}
\end{equation}
where the diagonal operator $D_{l}^{k_{1}k_{2}...k_{m}}$ $(1\leq m\leq n,$ $%
0\leq l\leq 2^{m}-1)$ is a basis operator of the operator set $\tilde{Q}_{B}$
given by definition (9). In particular, $C_{m,l}(\theta )\equiv
C_{l}^{12...m}(\theta ).$ The $m-$qubit selective rotation operation $%
C_{0}^{k_{1}k_{2}...k_{m}}(\theta )$ is only applied to any state $%
|0_{k_{1}}\rangle ...|0_{k_{2}}\rangle ...|0_{k_{m}}\rangle $ with the known
state $|0\rangle $ of the $k_{1}$th, $k_{2}$th, ..., $k_{m}$th spins of the $%
n-$qubit spin system. The diagonal operator $D_{0}^{k_{1}k_{2}...k_{m}}$ can
be really converted efficiently into the diagonal operator $D_{m,0}.$ This
can be achieved with the help of the zero-quantum unitary transformation: 
\begin{equation}
I_{kz}=V_{kl}(\pi )I_{lz}V_{kl}(\pi )^{+},  \label{13}
\end{equation}
where the zero-quantum unitary operator is given by 
\[
V_{kl}(\theta )=\exp [-i\theta \frac{1}{2}(2I_{kx}I_{ly}-2I_{ky}I_{lx})], 
\]
which can be decomposed into a sequence of two two-qubit quantum gates: 
\begin{equation}
V_{kl}(\theta )=\exp (-i\theta I_{kx}I_{ly})\exp (i\theta I_{ky}I_{lx}).
\label{14}
\end{equation}
Assume that the indices in the diagonal operator $D_{0}^{k_{1}k_{2}...k_{m}}$
satisfy $1\leq k_{1}<k_{2}<...<k_{m}\leq n$. If the index $k_{1}\neq 1$ then
making the zero-quantum unitary transformation (13) with the zero-quantum
unitary operator $V_{kl}(\pi )$ (14) with the indices $k=1$ and $l=k_{1}$ on
the diagonal operator $D_{0}^{k_{1}k_{2}...k_{m}}$ the operator $%
D_{0}^{k_{1}k_{2}...k_{m}}$ is converted into $D_{0}^{1k_{2}...k_{m}},$ and
if $k_{1}=1$ then $D_{0}^{k_{1}k_{2}...k_{m}}=D_{0}^{1k_{2}...k_{m}}$ and no
unitary transformation is needed to apply on the operator $%
D_{0}^{k_{1}k_{2}...k_{m}}.$ In a similar way, if the index $k_{2}\neq 2$
then $D_{0}^{1k_{2}...k_{m}}$ is converted into $D_{0}^{12k_{3}...k_{m}}$ by 
$V_{kl}(\pi )$ with the indices $k=2$ and $l=k_{2}$. By making the
zero-quantum unitary transformation (13) on the diagonal operator $%
D_{0}^{k_{1}k_{2}...k_{m}}$ at most $m$ times one can convert completely the
diagonal operator $D_{0}^{k_{1}k_{2}...k_{m}}$ into the diagonal operator $%
D_{0}^{12...m}$ $(\equiv D_{m,0}).$ Then the selective rotation operations $%
C_{0}^{k_{1}k_{2}...k_{m}}(\theta )$ (12) can be implemented efficiently
since $C_{m,0}(\theta )$ can be performed efficiently. Finally, using $m$
pulses $\exp (-i\pi I_{kx})$ or $\exp (-i\pi I_{ky})$ at most one can
convert the diagonal operator $D_{m,l}$ into $D_{m,0}$ and $%
D_{l}^{k_{1}k_{2}...k_{m}}$ into $D_{0}^{k_{1}k_{2}...k_{m}}$ for any index $%
l\neq 0$. This is based on the fact that $(\frac{1}{2}E_{k}+I_{kz})=\exp
(-i\pi I_{k\mu })(\frac{1}{2}E_{k}-I_{kz})\exp (i\pi I_{k\mu })$ $(\mu
=x,y). $ Then any $m-$qubit selective rotation operation $%
C_{l}^{k_{1}k_{2}...k_{m}}(\theta )$ (12) can be performed efficiently.

Therefore, the basic quantum gates used to construct quantum circuits for
multiple-quantum unitary operators include three types:\newline
$(i)$ the $m-$qubit selective rotation operations: $C_{m,0}(\theta )$ $%
(m=1,2,...,n),$\newline
$(ii)$ the two-qubit quantum gates: $\exp (-iJ_{kl}2I_{kz}I_{lz})$ $%
(k,l=1,2,...,n),$\newline
$(iii)$ the single-qubit gates: $\exp (-i\theta _{k\mu }I_{k\mu })$ $(\mu
=x,y,z;$ $k=1,2,...,n).$\newline
Obviously, the three types of basic quantum gates form the universal quantum
gate set in quantum computation.

Once the elementary propagators formed from the basis operators of the
product operator sets $\tilde{Q}_{A}$ and $\tilde{Q}_{B}$ can be implemented
efficiently then these basic unitary operations $U_{\beta }(\theta )=\exp
(-i\theta Q_{\beta })$ ($\beta =a,b,c$) built up with the following three
types of the basic product operators also can be efficiently implemented: 
\newline
$(i)$ $Q_{a}=2^{l-1}I_{k_{1}\mu _{1}}\bigotimes I_{k_{2}\mu _{2}}\bigotimes
...\bigotimes I_{k_{l}\mu _{l}}\bigotimes E_{k_{l+1}}\bigotimes
...\bigotimes E_{k_{n}};$ \newline
$(ii)$ $Q_{b}=(\frac{1}{2}E_{k_{1}}+a_{k_{1}}^{t}I_{k_{1}\mu
_{1}})\bigotimes (\frac{1}{2}E_{k_{2}}+a_{k_{2}}^{t}I_{k_{2}\mu
_{2}})\bigotimes ...$

$\bigotimes (\frac{1}{2}E_{k_{l}}+a_{k_{l}}^{t}I_{k_{l}\mu _{l}})\bigotimes
E_{k_{l+1}}\bigotimes ...\bigotimes E_{k_{n}};$ \newline
$(iii)$ $Q_{c}=2^{l-1}I_{k_{1}\mu _{1}}\bigotimes ...\bigotimes I_{k_{l}\mu
_{l}}\bigotimes (\frac{1}{2}E_{k_{l+1}}+a_{k_{l+1}}^{t}I_{k_{l+1}\mu
_{l+1}})\bigotimes ...$

$\bigotimes (\frac{1}{2}E_{k_{m}}+a_{k_{m}}^{t}I_{k_{m}\mu _{m}})\bigotimes
E_{k_{m+1}}\bigotimes ...\bigotimes E_{k_{n}},$ \newline
where $\mu _{\alpha }=x,y,z$ ($\alpha =1,2,...,n$)$.$ This is because each
of these tension product operators always can be efficiently converted into
a basis operator or a sum of two basis operators of the product operator
sets $\tilde{Q}_{A}$ and $\tilde{Q}_{B}$ with the help of the recursive
relation (10), the zero-quantum unitary transformation (13), and the
single-spin rotation operations: $\exp (-i\theta I_{k\mu }),\mu =x,y,z.$ As
an example, suppose that a product operator $Q_{c}$ is given by\newline
\[
Q_{c}=2I_{1x}\bigotimes E_{2}\bigotimes (\frac{1}{2}E_{3}+I_{3x})\bigotimes
I_{4z}\bigotimes (\frac{1}{2}E_{5}-I_{5z}). 
\]
Then it can be expressed as 
\begin{eqnarray*}
Q_{c} &=&R_{0}[(\frac{1}{2}E_{1}+I_{1z})\bigotimes E_{2}\bigotimes (\frac{1}{%
2}E_{3}+I_{3z})\bigotimes E_{4}\bigotimes (\frac{1}{2}E_{5}+I_{5z}) \\
&&-\frac{1}{2}E_{1}\bigotimes E_{2}\bigotimes (\frac{1}{2}%
E_{3}+I_{3z})\bigotimes E_{4}\bigotimes (\frac{1}{2}E_{5}+I_{5z})]R_{0}^{+}
\end{eqnarray*}
where the unitary operator $R_{0}$ is determined by the recursive relation
(10) and the single-spin rotation operations, 
\begin{eqnarray*}
R_{0} &=&\exp (i\frac{\pi }{2}I_{1y})\exp (i\frac{\pi }{2}I_{3y})\exp (-i\pi
I_{5x}) \\
&&\times \exp (-i\frac{\pi }{2}I_{1x})\exp (-i\pi I_{1z}I_{4z})\exp (-i\frac{%
\pi }{2}I_{1y}).
\end{eqnarray*}
If the zero-quantum unitary transformation (13) is further used then the
unitary operator $\exp (-i\theta Q_{c})$ can be thoroughly decomposed as a
sequence of the basic quantum gates: 
\begin{eqnarray*}
\exp (-i\theta Q_{c}) &=&R_{0}V_{23}(\pi )^{+}V_{35}(\pi )^{+}C_{3,0}(\theta
)V_{35}(\pi )V_{23}(\pi ) \\
&&\times V_{13}(\pi )^{+}V_{25}(\pi )^{+}C_{2,0}(-\theta /2)V_{13}(\pi
)V_{25}(\pi )R_{0}^{+}.
\end{eqnarray*}
\newline
\newline
{\large 3. The quantum circuits for the subspace-selective multiple-}

{\large quantum unitary operations}\newline
\newline
{\large 3.1 The Hermitian diagonal operators}\newline

Two types of Hermitian operators will be used to generate the
multiple-quantum unitary operators (5). The first type simply consists of
the $(n+1)$ Hermitian diagonal operators $\{g_{0},$ $g_{1},$ $g_{2},$ $...,$ 
$g_{n}\}$ for an $n-$qubit spin system. Each of the diagonal operators
corresponds one-to-one to one state subspace $S_{zq}(k).$ The diagonal
operators are generated from the quantum-state diagonal operator set $%
\{D_{k}\}$. Their definition is given below.

$g_{0}=D_{0},$ $D_{0}\in S_{zq}(0)\times S_{zq}(0),$\newline
or in the matrix representation the diagonal operator $g_{0}$ is written as%
\newline

$g_{0}=\left[ 
\begin{array}{llll}
1 &  &  &  \\ 
& 0 &  &  \\ 
&  & \ddots &  \\ 
&  &  & 0
\end{array}
\right] \equiv Diag(1_{0},0_{1},...,0_{N-1});$

$g_{1}=\stackrel{d(1)}{\stackunder{k}{\sum }}D_{k},$ $D_{k}\in
S_{zq}(1)\times S_{zq}(1),$\newline
or $g_{1}=\left[ 
\begin{array}{lllllll}
0 &  &  &  &  &  &  \\ 
& 1 &  &  &  &  &  \\ 
&  & \ddots &  &  &  &  \\ 
&  &  & 1 &  &  &  \\ 
&  &  &  & 0 &  &  \\ 
&  &  &  &  & \ddots &  \\ 
&  &  &  &  &  & 0
\end{array}
\right] $

$\equiv Diag(0_{0},1_{l_{1}},...,1_{L_{1}},0_{L_{1}+1},...,0_{N-1}),$

$l_{1}=d(0)=1,$ $L_{1}=l_{1}+d(1)-1;$

$g_{2}=\stackrel{d(2)}{\stackunder{k}{\sum }}D_{k},$ $D_{k}\in
S_{zq}(2)\times S_{zq}(2),$\newline
or $%
g_{2}=Diag(0_{0},...,0_{l_{2}-1},1_{l_{2}},...,1_{L_{2}},0_{L_{2}+1},...,0_{N-1}), 
$

$l_{2}=d(0)+d(1),$ $L_{2}=l_{2}+d(2)-1;$

$......;$

$g_{m}=\stackrel{d(m)}{\stackunder{k}{\sum }}D_{k},$ $D_{k}\in
S_{zq}(m)\times S_{zq}(m),$\newline
or $%
g_{m}=Diag(0_{0},...,0_{l_{m}-1},1_{l_{m}},...,1_{L_{m}},0_{L_{m}+1},...,0_{N-1}), 
$

$l_{m}=d(0)+d(1)+...+d(m-1),L_{m}=l_{m}+d(m)-1;$

$......;$

$g_{n}=D_{N-1},$ $D_{N-1}\in S_{zq}(n)\times S_{zq}(n),$\newline
or

$g_{n}=\left[ 
\begin{array}{llll}
0 &  &  &  \\ 
& \ddots &  &  \\ 
&  & 0 &  \\ 
&  &  & 1
\end{array}
\right] \equiv Diag(0_{0},0_{1},...,0_{N-2},1_{N-1}),$\newline
\newline
where again $d(m)=\binom{n}{m}$ is dimensional size of the subspace $%
S_{zq}(m)$; $l_{m}$ and $L_{m}$ are diagonal-element indices in the matrix $%
g_{m}$; and $S_{zq}(m)\times S_{zq}(m)$ is the zero-quantum operator
subspace corresponding to the subspace $S_{zq}(m)$ whose basis operator set
may be simply $\{|k\rangle \langle l|\}$ with any basis states $|k\rangle
,|l\rangle \in S_{zq}(m)$. This zero-quantum operator subspace also includes
the diagonal operator $g_{m}.$ Obviously, the unitary rotation operation $%
G_{m}(\theta )=\exp (-i\theta g_{m})$ $(m=0,1,2,...,n)$ can be expressed as
a sequence of the selective rotation operations applied only on the states
of the subspace $S_{zq}(m)$: 
\begin{equation}
G_{m}(\theta )=\stackrel{L_{m}}{\stackunder{k=l_{m}}{\prod }}C_{k}(\theta )
\label{15}
\end{equation}
where $C_{k}(\theta )\equiv C_{n,k}(\theta )=\exp (-i\theta D_{k})$ given by
(11)\ with the diagonal operator $D_{k}\in S_{zq}(m)\times S_{zq}(m)$ is a
selective rotation operation applied only to the computational base $%
|\varphi _{k}\rangle $ of the subspace $S_{zq}(m)$, where the computational
base $|\varphi _{k}\rangle =\stackrel{n}{\stackunder{l=1}{\bigotimes }}(%
\frac{1}{2}T_{l}+a_{l}^{k}S_{l}),$ $T_{l}=|0_{l}\rangle +|1_{l}\rangle $ and 
$S_{l}=\frac{1}{2}(|0_{l}\rangle -|1_{l}\rangle )$ [2]$.$ It can be seen
from (15) that the rotation operation $G_{m}(\theta )$ of the subspace $%
S_{zq}(m)\times S_{zq}(m)$ with $m\sim n/2$ is an exponential sequence of
the selective rotation operations with number $d(m)$. But it may be really
simplified and its efficient quantum circuit may be built up with the
elementary propagators $R_{k_{1}k_{2}...k_{m}}(\theta )$ (10) and $%
C_{l}^{k_{1}k_{2}...k_{l}}(\theta )$ (12).

The diagonal matrix $g_{m}$ has $d(m)$ and $(2^{n}-d(m))$ diagonal elements
taking one and zero, respectively. The $(2^{n}-d(m))$ zero-diagonal elements
are divided into two parts by the $d(m)$ one-diagonal elements, and numbers
for the first part and the second are $l_{m}$ and $(2^{n}-d(m)-l_{m}),$
respectively. Note that $2^{n}=l_{m}+d(m)+(2^{n}-d(m)-l_{m})$ is an even
number. Then there are only two possibilities: (i) all the three numbers $%
l_{m},$ $d(m),$ and $(2^{n}-d(m)-l_{m})$ are even or (ii) one of the three
numbers must be even and the other two numbers are odd. For the first case
that all the three numbers are even the diagonal operator $g_{m}$ can be
reduced to the form 
\begin{eqnarray*}
g_{m} &\equiv &Diag(0,...,0;1_{l_{m}},...,1_{L_{m}};0,...,0) \\
&=&Diag(0,...,0;1_{l_{m}/2},...,1_{L_{m}/2};0,...,0)_{2^{n-1}\times
2^{n-1}}\bigotimes E_{n}.
\end{eqnarray*}
Now the new diagonal operator $Diag(0,$ $...,$ $0;$ $%
1_{l_{m}/2},...,1_{L_{m}/2};$ $0,...,0)_{2^{n-1}\times 2^{n-1}}$ is applied
only to the subsystem with the first $n-1$ qubits of the $n-$qubit spin
system. Its dimensional size is $2^{n-1}\times 2^{n-1},$ which is denoted in
the subscript for convenience, instead of dimensional size $2^{n}\times
2^{n} $ of the original diagonal operator $g_{m}$. The number of the
diagonal elements taking one in the operator $%
Diag(0,...,0;1,...,1;0,...,0)_{2^{n-1}\times 2^{n-1}}$ now is $d(m)/2$. The
unitary operation $G_{m}(\theta )$ then is simplified (as defined), 
\begin{eqnarray*}
G_{m}(\theta ) &=&\exp [-i\theta
Diag(0,...,0;1_{l_{m}/2},...,1_{L_{m}/2};0,...,0)_{2^{n-1}\times
2^{n-1}}\bigotimes E_{n}] \\
&\equiv &\exp [-i\theta
Diag(0,...,0;1_{l_{m}/2},...,1_{L_{m}/2};0,...,0)_{2^{n-1}\times 2^{n-1}}].
\end{eqnarray*}
Thus, the unitary operation $G_{m}(\theta )$ is really applied only to the
subsystem with the first $n-1$ qubits of the $n-$qubit spin system and hence
is reduced to a $(n-1)-$qubit rotation operation of the $(n-1)-$qubit spin
subsystem. If now the rotation operation $G_{m}(\theta )$ is still expressed
as a sequence of the selective rotation operation applied to the $(n-1)-$%
qubit subsystem then number of the selective rotation operations, i.e., $%
d(m)/2,$ in the sequence is a half of the original one. Obviously, such a
reduction can be further carried out.

It is slightly complicated for another case that only one number is even
among the three numbers $l_{m},$ $d(m),$ and $(2^{n}-d(m)-l_{m})$. In this
case there are only three possible options:

(a) $l_{m}$ is an even number. Then $d(m)$ is an odd number. The diagonal
operator $g_{m}$ can be simplified by 
\[
g_{m}=Diag(0,...,0;1_{l_{m}/2},...,1_{(L_{m}-1)/2};0,...,0)_{2^{n-1}\times
2^{n-1}}\bigotimes E_{n}+D_{L_{m}} 
\]
where the diagonal operator $%
D_{L_{m}}=Diag(0,...,0;0_{l_{m}},...,0_{L_{m}-1},1_{L_{m}};0,...,0),$ and
the diagonal operator $%
Diag(0,...,0;1_{l_{m}/2},...,1_{(L_{m}-1)/2};0,...,0)_{2^{n-1}\times
2^{n-1}} $ has $(d(m)-1)/2$ one-diagonal elements and is a $(n-1)-$qubit
diagonal operator. Because all the diagonal operators are commutative to
each other the corresponding unitary operation $G_{m}(\theta )$ is
decomposed as 
\[
G_{m}(\theta )=C_{L_{m}}(\theta )\exp [-i\theta
Diag(0,...,0;1_{l_{m}/2},...,1_{(L_{m}-1)/2};0,...,0)_{2^{n-1}\times
2^{n-1}}]. 
\]
Therefore, the diagonal unitary operation $G_{m}(\theta )$ is decomposed
into a product of an $n-$qubit selective rotation operation $%
C_{L_{m}}(\theta )$ and a $(n-1)-$qubit diagonal unitary operation of the $%
n- $qubit spin system.

(b) $d(m)$ is an even number. Then the diagonal operator $g_{m}$ is written
as 
\begin{eqnarray*}
g_{m} &=&D_{l_{m}}+D_{L_{m}} \\
&&+Diag(0,...,0;1_{(l_{m}+1)/2},...,1_{(L_{m}-1)/2};0,...,0)_{2^{n-1}\times
2^{n-1}}\bigotimes E_{n}.
\end{eqnarray*}
The diagonal operator $%
Diag(0,...,0;1_{(l_{m}+1)/2},...,1_{(L_{m}-1)/2};0,...,0)_{2^{n-1}\times
2^{n-1}}$ is a $(n-1)-$qubit diagonal operator with $(d(m)-2)/2$
one-diagonal elements. The corresponding unitary operation $G_{m}(\theta )$
is expressed as 
\begin{eqnarray*}
G_{m}(\theta ) &=&C_{l_{m}}(\theta )C_{L_{m}}(\theta ) \\
&&\times \exp [-i\theta
Diag(0,...,0;1_{(l_{m}+1)/2},...,1_{(L_{m}-1)/2};0,...,0)_{2^{n-1}\times
2^{n-1}}].
\end{eqnarray*}
This shows that the diagonal unitary operation $G_{m}(\theta )$ now is
decomposed into a product of two $n-$qubit selective rotation operations $%
C_{l_{m}}(\theta )$ and $C_{L_{m}}(\theta )$ and a $(n-1)-$qubit diagonal
unitary operation.

(c) $(2^{n}-d(m)-l_{m})$ is an even number. The diagonal operator $g_{m}$ is
simplified by 
\[
g_{m}=D_{l_{m}}+Diag(0,...,0;1_{(l_{m}+1)/2},...,1_{L_{m}/2};0,...,0)_{2^{n-1}\times 2^{n-1}}\bigotimes E_{n}. 
\]
The diagonal operator $%
Diag(0,...,0;1_{(l_{m}+1)/2},...,1_{L_{m}/2};0,...,0)_{2^{n-1}\times
2^{n-1}} $ now is a $(n-1)-$qubit diagonal operator with $(d(m)-1)/2$
one-diagonal elements. The corresponding unitary operation $G_{m}(\theta )$
can be decomposed as 
\[
G_{m}(\theta )=C_{l_{m}}(\theta )\exp [-i\theta
Diag(0,...,0;1_{(l_{m}+1)/2},...,1_{L_{m}/2};0,...,0)_{2^{n-1}\times
2^{n-1}}]. 
\]
Thus, the diagonal unitary operation $G_{m}(\theta )$ now is decomposed into
a product of one $n-$qubit selective rotation operation $C_{l_{m}}(\theta )$
and another $(n-1)-$qubit diagonal unitary operation.

As a summary, in either case the $n-$qubit diagonal unitary operator $%
G_{m}(\theta )$ can be reduced to a product of a $(n-1)-$qubit diagonal
unitary operator and two $n-$qubit selective rotation operators at most.

The $(n-1)-$qubit rotation operation $\exp [-i\theta Diag(0,...,$ $0;1$ $%
,...,$ $1;0$ $,...,0)_{2^{n-1}\times 2^{n-1}}]$ can be further reduced to
the $(n-2)-$qubit rotation operation $\exp [-i\theta Diag(0,...,0;$ $%
1,...,1; $ $0,...,0)_{2^{n-2}\times 2^{n-2}}]$ which has around $d(m)/2^{2}$
one-diagonal elements, but this reduction may yield extra two $(n-1)-$qubit
selective rotation operations $C_{n-1,t_{m}}(\theta )$ (the index $t_{m}$ is
dependent on $l_{m}$ and $L_{m}$) at most, so that the unitary operator $%
G_{m}(\theta )$ now is a product of a $(n-2)-$qubit rotation operation and
four $n-$ and $(n-1)-$qubit selective rotation operations at most. This
reduction process can continue to the end when the diagonal operator $g_{m}$
is reduced to the final form 
\begin{equation}
Diag(0,...,0;1,1;0,...,0)_{2^{n-k}\times 2^{n-k}}\bigotimes
E_{k+1}\bigotimes ...\bigotimes E_{n}  \label{16}
\end{equation}
or 
\begin{equation}
Diag(0,...,0;1;0,...,0)_{2^{n-k}\times 2^{n-k}}\bigotimes E_{k+1}\bigotimes
...\bigotimes E_{n},  \label{17}
\end{equation}
where $k$ satisfies $2^{k}\thickapprox d(m)$ and is less than $n-1$ because $%
d(m)<2^{n-1}$ for $n>2$. The first diagonal operator (16) can form two $%
(n-k)-$qubit selective rotation operations and the second (17) can generate
only one $(n-k)-$qubit selective rotation operation. Since each reduction
step can generate two selective rotation operations $C_{l,t_{m}}(\theta )$
at most the diagonal unitary operator $G_{m}(\theta )$ can be expressed as a
sequence of $l-$qubit selective rotation operations $C_{l,t_{m}}(\theta )$ $%
(l=k,$ $k+1,$ $...,$ $n)$ with a total number less than $2n$.

The same decomposition procedure as the above can be carried out for a
general diagonal operator $%
Diag(0_{0},...,0_{l-1};1_{l},...,1_{L};0_{L+1},...,0_{N-1})$ that may not be
in only one zero-quantum operator subspace, e.g., $S_{zq}(m)\times
S_{zq}(m), $ and consequently the diagonal operator may be expressed as a
sequence of $l-$qubit selective rotation operations $C_{l,t_{m}}(\theta )$ $%
(l=k,$ $k+1,$ $...,$ $n)$ with number less than $2k,$ where $k$ satisfies $%
2^{k}\thickapprox (L-l+1)$ and is less than $n$. \newline
\newline
{\large 3.2 The Hermitian anti-diagonal operators}

Another type of the Hermitian operators used to generate the
subspace-selective multiple-quantum unitary operators are anti-diagonal
Hermitian operators. They are a generalization of the product operator $%
2^{n-1}I_{1x}I_{2x}...I_{nx}$ [2, 6] and are defined in a matrix form by%
\newline

$b_{0}=\left[ 
\begin{array}{llllll}
&  &  &  & 0 & 1 \\ 
&  &  & 0 & 1 & 0 \\ 
&  &  & . & 0 &  \\ 
&  & . &  &  &  \\ 
0 & . &  &  &  &  \\ 
1 & 0 &  &  &  & 
\end{array}
\right] ,$

$b_{1}=\left[ 
\begin{array}{llllll}
&  &  & 0 & 1 & 0 \\ 
&  & 0 & 1 & 0 &  \\ 
&  & . & . &  &  \\ 
0 & . & . &  &  &  \\ 
1 & . &  &  &  &  \\ 
0 &  &  &  &  & 
\end{array}
\right] ,$ $b_{-1}=\left[ 
\begin{array}{llllll}
&  &  &  &  & 0 \\ 
&  &  &  & 0 & 1 \\ 
&  &  & . & 1 & 0 \\ 
&  & . & . & 0 &  \\ 
& . & . &  &  &  \\ 
0 & 1 & 0 &  &  & 
\end{array}
\right] ,$

$b_{2}=\left[ 
\begin{array}{lllllll}
&  &  & 0 & 1 & 0 & 0 \\ 
&  & 0 & 1 & 0 & 0 &  \\ 
&  & . & . & . &  &  \\ 
0 & . & . & . &  &  &  \\ 
1 & . & . &  &  &  &  \\ 
0 & . &  &  &  &  &  \\ 
0 &  &  &  &  &  & 
\end{array}
\right] ,$ $b_{-2}=\left[ 
\begin{array}{lllllll}
&  &  &  &  &  & 0 \\ 
&  &  &  &  & 0 & 0 \\ 
&  &  &  & . & 0 & 1 \\ 
&  &  & . & . & 1 & 0 \\ 
&  & . & . & . & 0 &  \\ 
& . & . & . &  &  &  \\ 
0 & 0 & 1 & 0 &  &  & 
\end{array}
\right] ,$ $...,$

$b_{k}=\left[ 
\begin{array}{llllllll}
&  &  & 0 & 1 & 0 & \cdots & 0 \\ 
&  & 0 & 1 & 0 & \cdots & 0 &  \\ 
&  & . & . &  & . &  &  \\ 
0 & . & . &  & . &  &  &  \\ 
1 & . &  & . &  &  &  &  \\ 
0 &  & . &  &  &  &  &  \\ 
\vdots & . &  &  &  &  &  &  \\ 
0 &  &  &  &  &  &  & 
\end{array}
\right] ,$ $b_{-k}=\left[ 
\begin{array}{llllllll}
&  &  &  &  &  &  & 0 \\ 
&  &  &  &  &  & 0 & \vdots \\ 
&  &  &  &  & . & \vdots & 0 \\ 
&  &  &  & . &  & 0 & 1 \\ 
&  &  & . &  & . & 1 & 0 \\ 
&  & . &  & . & . & 0 &  \\ 
& . &  & . & . &  &  &  \\ 
0 & \cdots & 0 & 1 & 0 &  &  & 
\end{array}
\right] .$\newline
\newline
There are $2(2^{n}-2)+1$ anti-diagonal operators $b_{k}$ $(k=0,$ $\pm 1,$ $%
\pm 2,$ $...,$ $\pm (2^{n}-2))$ in an $n-$qubit spin system except the two
operators $|0\rangle \langle 0|$ and $|2^{n}-1\rangle \langle 2^{n}-1|.$ The
two operators are anti-diagonal operators and also diagonal operators. They
are usually assigned to diagonal operators. All these anti-diagonal
operators are symmetrical and Hermitian operators. For every anti-diagonal
matrix $b_{k}$ all its non-zero elements that take one are located along an
anti-diagonal line of the matrix. The matrix $b_{0}$ is the main
anti-diagonal matrix where all its $2^{n}$ non-zero elements taking one are
located along the main anti-diagonal line. The two end points (row, column)
of the main anti-diagonal line are $(0,2^{n}-1)$ and $(2^{n}-1,0)$ in the
matrix $b_{0}$, respectively. There are only $(2^{n}-k)$ nonzero elements in
the matrix $b_{k}$ (or $b_{-k}$) $(2^{n}-2\geq k\geq 0)$ along the
anti-diagonal line of the matrix. Actually, there is a unit difference
between numbers of matrix element in two nearest anti-diagonal lines in a
matrix. Thus, nonzero-element number of the anti-diagonal matrix $b_{k+1}$
is one less than that of the matrix $b_{k}$, and since the matrix $b_{0}$
has $2^{n}$ nonzero elements the matrix $b_{k}$ has $(2^{n}-k)$ nonzero
elements. The two end points of the anti-diagonal line for the matrix $b_{k}$
are $(0,2^{n}-1-k)$ and $(2^{n}-1-k,0),$ respectively, and for the matrix $%
b_{-k}$ are $(k,2^{n}-1)$ and $(2^{n}-1,k),$ respectively. Denote that $x$
is row coordinate and $y$ column coordinate. Then the main anti-diagonal
line is given by $x=-y+2^{n}-1,$ and the anti-diagonal lines of the matrices 
$b_{k}$ and $b_{-k}$ are given by $x=-y+2^{n}-1-k$ and $x=-y+2^{n}-1+k,$
respectively. Note that an anti-diagonal operator $b_{k}$ is symmetric and
it has $(2^{n}-k)$ nonzero (one) elements. If the index $k$ is odd then the
matrix $b_{k}$ must contain a diagonal element taking one. This diagonal
element exactly locates at the position ($2^{n-1}-(k+1)/2,2^{n-1}-(k+1)/2$)
along same anti-diagonal line of the matrix $b_{k}$. However, there is not
any diagonal element in the anti-diagonal matrix $b_{k}$ with an even index $%
k$.

The unitary operator $B_{k}(\theta )=\exp (-i\theta b_{k})$ built up with
any Hermitian anti-diagonal operator $b_{k}$ by exponential mapping may be
decomposed into a sequence of the basic unitary operations. This may be
achieved by expressing the anti-diagonal operator as a sum of basic product
operators. Several important unitary operators $B_{k}(\theta )$ are given
below with their explicit decomposition. Note that an anti-diagonal operator 
$b_{\pm k}$ with $k\geq 2^{n-1}$ first can be simplified by 
\[
(b_{\pm k})_{2^{n}\times 2^{n}}=(\frac{1}{2}E_{1}\pm I_{1z})\bigotimes
(b_{\pm (k-2^{n-1})})_{2^{n-1}\times 2^{n-1}} 
\]
where $b_{\pm k}\equiv (b_{\pm k})_{2^{n}\times 2^{n}}$ with the subscript $%
2^{n}\times 2^{n}$ indicating dimensional size of the matrix $b_{k}$. Now
the index $(k-2^{n-1})<2^{n-1}$ and one needs to express only the $(n-1)-$%
qubit anti-diagonal operator $(b_{\pm (k-2^{n-1})})_{2^{n-1}\times 2^{n-1}}$
in the product operator form. Therefore, explicit product operator
expressions will be given only for those operators $b_{\pm k}$ with $%
k<2^{n-1}$ below.

(a) The main anti-diagonal operator $b_{0}.$ The operator $b_{0}$ can be
easily expressed as 
\begin{equation}
b_{0}=2^{n}I_{1x}\bigotimes I_{2x}\bigotimes ...\bigotimes I_{nx}.
\label{18}
\end{equation}
This is a simple anti-diagonal operator often used in previous papers [2, 3,
5, 6]. With the help of the single-spin rotation operations and the
recursive relation (10) the unitary operator $B_{0}(\theta )=\exp (-i\theta
b_{0})$ can be easily decomposed into an efficient sequence of one- and
two-qubit gates.

(b) The operator $b_{\pm 1}.$ The product operator expression for the
operator $b_{1}$ is slightly complicated. There is a recursive relation for
the anti-diagonal operator $b_{1}$:

$b_{1}\equiv (b_{1})_{2^{n}\times
2^{n}}=2^{n-1}I_{1x}I_{2x}...I_{n-1x}\bigotimes (\frac{1}{2}E_{n}+I_{nz})$

$+(b_{1})_{2^{n-1}\times 2^{n-1}}\bigotimes (\frac{1}{2}E_{n}-I_{nz}).$%
\newline
Using this recursive relation one can express the operator $b_{1}$ as a sum
of $n$ commutative product operators:

\[
b_{1}=(\frac{1}{2}E_{1}+I_{1z})\bigotimes (\frac{1}{2}E_{2}-I_{2z})%
\bigotimes ...\bigotimes (\frac{1}{2}E_{n}-I_{nz}) 
\]
\[
+2I_{1x}\bigotimes (\frac{1}{2}E_{2}+I_{2z})\bigotimes (\frac{1}{2}%
E_{3}-I_{3z})\bigotimes ...\bigotimes (\frac{1}{2}E_{n}-I_{nz}) 
\]
\[
+2^{2}I_{1x}I_{2x}\bigotimes (\frac{1}{2}E_{3}+I_{3z})\bigotimes (\frac{1}{2}%
E_{4}-I_{4z})\bigotimes ...\bigotimes (\frac{1}{2}E_{n}-I_{nz})+...... 
\]

\[
+2^{n-2}I_{1x}\bigotimes I_{2x}\bigotimes ...\bigotimes I_{n-2x}\bigotimes (%
\frac{1}{2}E_{n-1}+I_{n-1z})\bigotimes (\frac{1}{2}E_{n}-I_{nz}) 
\]
\begin{equation}
+2^{n-1}I_{1x}\bigotimes I_{2x}\bigotimes ...\bigotimes I_{n-1x}\bigotimes (%
\frac{1}{2}E_{n}+I_{nz}).  \label{19}
\end{equation}
Since all these product operators in the operator $b_{1}$ are commutative
the corresponding unitary operator $B_{1}(\theta )$ is decomposed as a
sequence of $n$ basic unitary operations, 
\begin{eqnarray}
B_{1}(\theta ) &=&\exp [-i\theta (\frac{1}{2}E_{1}+I_{1z})\bigotimes (\frac{1%
}{2}E_{2}-I_{2z})\bigotimes ...\bigotimes (\frac{1}{2}E_{n}-I_{nz})] 
\nonumber \\
&&\times \exp [-i\theta 2I_{1x}\bigotimes (\frac{1}{2}E_{2}+I_{2z})%
\bigotimes (\frac{1}{2}E_{3}-I_{3z})\bigotimes ...  \nonumber \\
&&\bigotimes (\frac{1}{2}E_{n}-I_{nz})]  \nonumber \\
&&\times \exp [-i\theta 2^{2}I_{1x}I_{2x}\bigotimes (\frac{1}{2}%
E_{3}+I_{3z})\bigotimes (\frac{1}{2}E_{4}-I_{4z})\bigotimes ...  \nonumber \\
&&\bigotimes (\frac{1}{2}E_{n}-I_{nz})]\times ......  \nonumber \\
&&\times \exp [-i\theta 2^{n-2}I_{1x}I_{2x}...I_{n-2x}\bigotimes (\frac{1}{2}%
E_{n-1}+I_{n-1z})\bigotimes (\frac{1}{2}E_{n}-I_{nz})]  \nonumber \\
&&\times \exp [-i\theta 2^{n-1}I_{1x}I_{2x}...I_{n-1x}\bigotimes (\frac{1}{2}%
E_{n}+I_{nz})].  \label{20}
\end{eqnarray}
In a similar way, the anti-diagonal operator $b_{-1}$ also can be expressed
as a sum of $n$ commutative product operators,

$b_{-1}=(\frac{1}{2}E_{1}-I_{1z})\bigotimes (\frac{1}{2}E_{2}+I_{2z})%
\bigotimes ...\bigotimes (\frac{1}{2}E_{n}+I_{nz})$

$+2I_{1x}\bigotimes (\frac{1}{2}E_{2}-I_{2z})\bigotimes (\frac{1}{2}%
E_{3}+I_{3z})\bigotimes ...\bigotimes (\frac{1}{2}E_{n}+I_{nz})$

$+2^{2}I_{1x}I_{2x}\bigotimes (\frac{1}{2}E_{3}-I_{3z})\bigotimes (\frac{1}{2%
}E_{4}+I_{4z})\bigotimes ...\bigotimes (\frac{1}{2}E_{n}+I_{nz})+......$

$+2^{n-2}I_{1x}\bigotimes I_{2x}\bigotimes ...\bigotimes I_{n-2x}\bigotimes (%
\frac{1}{2}E_{n-1}-I_{n-1z})\bigotimes (\frac{1}{2}E_{n}+I_{nz})$

$+2^{n-1}I_{1x}\bigotimes I_{2x}\bigotimes ...\bigotimes I_{n-1x}\bigotimes (%
\frac{1}{2}E_{n}-I_{nz}),$\newline
and corresponding unitary operator $B_{-1}(\theta )$ therefore is decomposed
as a sequence of $n$ basic unitary operations.

(c) The operator $b_{\pm k}$ with $k=2^{l}$ $(l=1,2,...,n-1).$ First
consider the operator $b_{k}$ with an even index $k.$ The number of nonzero
(one) elements of the matrix $b_{k}$ along the anti-diagonal line is $%
2^{n}-k.$ If the index $k$ is even then so is $2^{n}-k$. If now the matrix $%
b_{k}$ is blocked by a $2\times 2$ submatrix one can see this blocked matrix 
$b_{k}$ is still an anti-diagonal blocked matrix, and the nonzero blocked
submatrix is $2I_{x}=\left( 
\begin{array}{ll}
0 & 1 \\ 
1 & 0
\end{array}
\right) $ along the anti-diagonal line. Therefore, the matrix $b_{k}$ may be
written as $(b_{k})_{2^{n}\times 2^{n}}=(b_{k^{\prime }})_{2^{n-1}\times
2^{n-1}}\bigotimes 2I_{nx}$ with index $k^{\prime }=k/2,$ and $b_{k^{\prime
}}$ is also an anti-diagonal matrix. If $k^{\prime }$ is still even then the
matrix $b_{k}$ can be further written as $(b_{k})_{2^{n}\times
2^{n}}=(b_{k^{\prime \prime }})_{2^{n-2}\times 2^{n-2}}\bigotimes
2I_{n-1x}\bigotimes 2I_{nx}$ with index $k^{\prime \prime }=k/4$. Generally,
for the operator $b_{k}$ with $k=2^{l}$ $(l=1,2,...,n-1)$ number of nonzero
elements is $(2^{n}-2^{l})$ on the anti-diagonal line. Then the operator $%
b_{2^{l}}$ can be reduced to the form 
\begin{equation}
b_{2^{l}}=(b_{1})_{2^{n-l}\times 2^{n-l}}\bigotimes
2^{l}I_{n-l+1x}I_{n-l+2x}...I_{nx}.  \label{21}
\end{equation}
Particularly the matrix $b_{2}$ can be written as $b_{2}=(b_{1})_{2^{n-1}%
\times 2^{n-1}}\bigotimes 2I_{nx}.$ Because the operator $%
(b_{1})_{2^{n-l}\times 2^{n-l}}$ can be further expressed as a sum of $(n-l)$
commutative product operators, as shown in (b), the operator $b_{2^{l}}$ now
is written as a sum of $(n-l)$ commutable product operators and hence the
unitary operator $B_{k}(\theta )=\exp (-i\theta b_{k})$ with $k=2^{l}$ $%
(l=1,2,...,n-1)$ can be decomposed as a sequence of $(n-l)$ basic unitary
operations. In a similar way, the anti-diagonal operator $b_{-k}$ with $%
k=2^{l}$ can be reduced to the form 
\[
b_{-2^{l}}=(b_{-1})_{2^{n-l}\times 2^{n-l}}\bigotimes
2^{l}I_{n-l+1x}I_{n-l+2x}...I_{nx}, 
\]
where $(b_{-1})_{2^{n-l}\times 2^{n-l}}$ can be further expressed as a sum
of $(n-l)$ commutative product operators. Therefore, the unitary operator $%
B_{-k}(\theta )=\exp (-i\theta b_{-k})$ with $k=2^{l}$ can also be
decomposed as a sequence of $(n-l)$ basic unitary operations.

Generally, an anti-diagonal operator $b_{k}$ with an even index $%
k=2^{k_{l-1}}+2^{k_{l-2}}+...+2^{k_{1}}$ can be simplified by 
\begin{equation}
b_{k}=(b_{k^{\prime }})_{2^{n-k_{1}}\times 2^{n-k_{1}}}\bigotimes
2^{k_{1}}I_{n-k_{1}+1x}I_{n-k_{1}+2x}...I_{nx}  \label{22}
\end{equation}
with the odd index $k^{\prime
}=2^{k_{l-1}-k_{1}}+2^{k_{l-2}-k_{1}}+...+2^{k_{2}-k_{1}}+1.$

(d) A general operator $b_{k}$. In (c) it is shown that an anti-diagonal
operator $b_{k}$ with an even index $k$ can be reduced to another
lower-dimensional anti-diagonal operator with an odd index. Here consider
the operator $b_{k}$ with an odd index $k$. The operator $b_{k}$ always can
be written in the form\newline
\begin{equation}
(b_{k})_{2^{n}\times 2^{n}}=\left\{ 
\begin{array}{l}
(b_{k_{1}/2})_{2^{n-2}\times 2^{n-2}}\bigotimes 2I_{n-1x}\bigotimes (\frac{1%
}{2}E_{n}+I_{nz}) \\ 
\quad +(b_{l_{1}})_{2^{n-1}\times 2^{n-1}}\bigotimes (\frac{1}{2}%
E_{n}-I_{nz}),\text{ if }k_{1}\text{ is even,} \\ 
(b_{l_{1}/2})_{2^{n-2}\times 2^{n-2}}\bigotimes 2I_{n-1x}\bigotimes (\frac{1%
}{2}E_{n}-I_{nz}) \\ 
\quad +(b_{k_{1}})_{2^{n-1}\times 2^{n-1}}\bigotimes (\frac{1}{2}%
E_{n}+I_{nz}),\text{ if }l_{1}\text{ is even,}
\end{array}
\right. \newline
\label{23}
\end{equation}
where $k_{1}=(k-1)/2$ and $l_{1}=(k+1)/2.$ The number of nonzero elements
keep unchanged before and after the reduction: $%
(2^{n}-k)=(2^{(n-1)}-k_{1})+(2^{(n-1)}-l_{1}).$ The $n-$qubit anti-diagonal
operator $(b_{k})_{2^{n}\times 2^{n}}$ now consists of one term containing $%
(n-1)-$qubit anti-diagonal operator $(b_{l_{1}})_{2^{n-1}\times 2^{n-1}}$
(or $(b_{k_{1}})_{2^{n-1}\times 2^{n-1}})$ and another term containing $%
(n-2)-$qubit anti-diagonal operator $(b_{k_{1}/2})_{2^{n-2}\times 2^{n-2}}$
(or $(b_{l_{1}/2})_{2^{n-2}\times 2^{n-2}})$. Note that the two terms are
commutable to each other. This is an advantage of the decomposition based on
the recursive relation (23). Image that using (23) an $n-$qubit
anti-diagonal operator at the first step is decomposed as a sum of two $%
(n-1)-$qubit anti-diagonal operator terms, then at the second step the two $%
(n-1)-$qubit anti-diagonal operators are decomposed as four $(n-2)-$qubit
anti-diagonal operator terms, and so on, in the final the $n-$qubit operator 
$b_{k}$ would be a sum of $2^{n-1}$ commutative basic product operators if
the decomposition could be carried out to the final $(n-1)$th step. However,
the recursive relation (23) shows that although the $n-$qubit operator $%
b_{k} $ is first decomposed as a sum of two $(n-1)-$qubit anti-diagonal
operator terms, one of the two $(n-1)-$qubit anti-diagonal operators can be
further reduced to only one $(n-2)-$qubit anti-diagonal operator term
instead of two terms. Then term number of the basic product operators in the
completely decomposed operator $b_{k}$ is not really more than $2^{n-1}$.
For some specific cases one may use conveniently the recursive relation (23)
to decompose an anti-diagonal operator $b_{k}$ as a sum of polynomially many
basic product operators. Take the anti-diagonal operators $b_{\pm k}$ with
index $k=2^{r}\pm 2^{m}$ $(n>$ $r>m=0,1,2,...,n-1)$ as an example$.$

For the index $k=2^{r}\pm 2^{m}$ the operator $b_{k}$ is first reduced to
the form $(b_{k})_{2^{n}\times 2^{n}}=(b_{2^{r-m}\pm 1})_{2^{n-m}\times
2^{n-m}}\bigotimes 2^{m}I_{n-m+1x}I_{n-m+2x}...I_{nx}$, as shown in (22)$.$
Then consider the anti-diagonal operator $(b_{2^{l}+1})_{2^{t}\times 2^{t}}$
with $t=n-m$ and $l=r-m.$ It follows from (23)\ that there is a recursive
relation for the operator $(b_{2^{l}+1})_{2^{t}\times 2^{t}}$: 
\begin{eqnarray*}
(b_{2^{l}+1})_{2^{t}\times 2^{t}} &=&(b_{2^{l-1}+1})_{2^{t-1}\times
2^{t-1}}\bigotimes (\frac{1}{2}E_{t}-I_{tz}) \\
&&+(b_{1})_{2^{t-l}\times 2^{t-l}}\bigotimes
2^{l-1}I_{t-l+1x}I_{t-l+2x}...I_{t-1x}\bigotimes (\frac{1}{2}E_{t}+I_{tz}).
\end{eqnarray*}
This is because in (23) $k_{1}=$ $(k-1)/2=2^{l-1}$ and $%
(b_{2^{l-1}/2})_{2^{t-2}\times 2^{t-2}}$ can be further reduced to the
operator $(b_{1})_{2^{t-l}\times 2^{t-l}},$ as shown in (c). This relation
directly leads to the product operator expression for the operator $%
(b_{2^{l}+1})_{2^{t}\times 2^{t}}:$

$(b_{2^{l}+1})_{2^{t}\times 2^{t}}=(b_{1})_{2^{t-l}\times 2^{t-l}}\bigotimes
2^{l-1}I_{t-l+1x}I_{t-l+2x}...I_{t-1x}\bigotimes (\frac{1}{2}E_{t}+I_{tz})$

$+(b_{1})_{2^{t-l}\times 2^{t-l}}\bigotimes
2^{l-2}I_{t-l+1x}I_{t-l+2x}...I_{t-2x}\bigotimes (\frac{1}{2}%
E_{t-1}+I_{t-1z})$

$\bigotimes (\frac{1}{2}E_{t}-I_{tz})$

$+(b_{1})_{2^{t-l}\times 2^{t-l}}\bigotimes
2^{l-3}I_{t-l+1x}I_{t-l+2x}...I_{t-3x}\bigotimes (\frac{1}{2}%
E_{t-2}+I_{t-2z})$

$\bigotimes (\frac{1}{2}E_{t-1}-I_{t-1z})\bigotimes (\frac{1}{2}%
E_{t}-I_{tz})+......$

$+(b_{1})_{2^{t-l}\times 2^{t-l}}\bigotimes 2I_{t-l+1x}\bigotimes (\frac{1}{2%
}E_{t-l+2}+I_{t-l+2z})\bigotimes (\frac{1}{2}E_{t-l+3}-I_{t-l+3z})$

$\bigotimes ...\bigotimes (\frac{1}{2}E_{t}-I_{tz})$

$+(b_{1})_{2^{t-l}\times 2^{t-l}}\bigotimes (\frac{1}{2}%
E_{t-l+1}+I_{t-l+1z})\bigotimes (\frac{1}{2}E_{t-l+2}-I_{t-l+2z})\bigotimes
...$

$\bigotimes (\frac{1}{2}E_{t}-I_{tz})$

$+(b_{2})_{2^{t-l}\times 2^{t-l}}\bigotimes (\frac{1}{2}%
E_{t-l+1}-I_{t-l+1z})\bigotimes (\frac{1}{2}E_{t-l+2}-I_{t-l+2z})\bigotimes
...$

$\bigotimes (\frac{1}{2}E_{t}-I_{tz}).$\newline
The operator $(b_{2^{l}+1})_{2^{t}\times 2^{t}}$ now is a sum of $(l+1)$
commutative operators. Since the operator $(b_{1})_{2^{t-l}\times 2^{t-l}}$
can be expressed as a sum of $(t-l)$ commutative product operators, as shown
in (b), and $(b_{2})_{2^{t-l}\times 2^{t-l}}$ as a sum of $(t-l-1)$
commutative product operators, as shown in (c), then the operator $%
(b_{2^{l}+1})_{2^{t}\times 2^{t}}$ is a sum of $(l+1)(t-l)-1$ commutative
product operators. Now taking $t=$ $n-m$ and $l=r-m$ one sees the operator $%
(b_{k})_{2^{n}\times 2^{n}}$ with $k=2^{r}+2^{m}$ is a sum of $%
(r-m+1)(n-r)-1 $ commutative product operators. Thus, the unitary operator $%
B_{k}(\theta )$ can be decomposed as a sequence of $(r-m+1)(n-r)-1$ basic
unitary operations. In an analogous way, it can be shown that the unitary
operator $B_{-k}(\theta )$ can be also decomposed as a sequence of $%
(r-m+1)(n-r)-1$ basic unitary operations.

From (23) there is also a recursive relation for the operator $b_{2^{l}-1}:$ 
\[
(b_{2^{l}-1})_{2^{t}\times 2^{t}}=(b_{2^{l-1}-1})_{2^{t-1}\times
2^{t-1}}\bigotimes (\frac{1}{2}E_{t}+I_{tz})+(b_{2^{l-1}})_{2^{t-1}\times
2^{t-1}}\bigotimes (\frac{1}{2}E_{t}-I_{tz}). 
\]
Then using this relation the product operator expression for the operator $%
(b_{2^{l}-1})_{2^{t}\times 2^{t}}$ is given by 
\begin{eqnarray*}
(b_{2^{l}-1})_{2^{t}\times 2^{t}} &=&(b_{2^{l-1}})_{2^{t-1}\times
2^{t-1}}\bigotimes (\frac{1}{2}E_{t}-I_{tz}) \\
&&+(b_{2^{l-2}})_{2^{t-2}\times 2^{t-2}}\bigotimes (\frac{1}{2}%
E_{t-1}-I_{t-1z})\bigotimes (\frac{1}{2}E_{t}+I_{tz}) \\
&&+......+(b_{2})_{2^{t-l+1}\times 2^{t-l+1}}\bigotimes (\frac{1}{2}%
E_{t-l+2}-I_{t-l+2z}) \\
&&\bigotimes (\frac{1}{2}E_{t-l+3}+I_{t-l+3z})\bigotimes ...\bigotimes (%
\frac{1}{2}E_{t}+I_{tz}) \\
&&+(b_{1})_{2^{t-l+1}\times 2^{t-l+1}}\bigotimes (\frac{1}{2}%
E_{t-l+2}+I_{t-l+2z}) \\
&&\bigotimes (\frac{1}{2}E_{t-l+3}+I_{t-l+3z})\bigotimes ...\bigotimes (%
\frac{1}{2}E_{t}+I_{tz}).
\end{eqnarray*}
Because the operator $(b_{2^{l-m}})_{2^{t-m}\times 2^{t-m}}$ $%
(m=1,2,...,l-1) $ can be expressed as a sum of $(t-l)$ commutative product
operators the operator $(b_{2^{l}-1})_{2^{t}\times 2^{t}}$ is clearly a sum
of $l(t-l)+1$ commutative product operators. Then the operator $b_{k}$ with $%
k=2^{r}-2^{m}$ is a sum of $(r-m)(n-r)+1$ commutative product operators and
hence its unitary operator $B_{k}(\theta )$ can be decomposed as a sequence
of $(r-m)(n-r)+1$ basic unitary operations.

The above results suffice to construct an efficient subspace- selective
multiple-quantum unitary operator (5) that transfers completely an unknown
state from any subspace $S_{zq}(m)$ ($m\neq n/2$) to the largest subspace $%
S_{zq}(n/2)$ of the Hilbert space of an $n-$qubit spin system.

A general anti-diagonal operator also can be decomposed using the recursive
relation (23) in an analogue way. But by using only the recursive relation
(23) it is usually not convenient to obtain an efficient decomposition for
the unitary operator $B_{k}(\theta )=\exp (-i\theta b_{k})$ with a general
anti-diagonal operator $b_{k}$. It can be proved that a general
anti-diagonal operator can be converted into another anti-diagonal operator
with a different index by a proper unitary transformation, and for a general
anti-diagonal operator $b_{k}$ with an odd index $%
k=2^{k_{l-1}}+2^{k_{l-2}}+...+2^{k_{1}}+1$ ($n-1>k_{l-1}>k_{l-2}>...>k_{1}%
\geq 1)$ its unitary operator $B_{k}(\theta )=\exp (-i\theta b_{k})$ can be
generally expressed as 
\begin{equation}
B_{k}(\theta )=U_{k}\exp (-i\theta \overline{b}_{k1})U_{k}^{+}  \label{24}
\end{equation}
where the operator $\overline{b}_{k1}$ is a symmetric and Hermitian
anti-diagonal operator similar to the anti-diagonal operator $b_{1}$ and $%
U_{k}$ is a unitary operator dependent on the index $k$. The anti-diagonal
line of the operator $\overline{b}_{k1}$ is the same as that of the operator 
$b_{1}$, but number of nonzero elements taking one along the anti-diagonal
line in the operator $\overline{b}_{k1}$ is only $(2^{n}-k)$ instead of $%
(2^{n}-1)$ of the operator $b_{1}.$ The $(2^{n}-k)$ nonzero elements locate
symmetrically at the center of the anti-diagonal line and each of two ends
of the anti-diagonal line has $(k-1)/2$ zero elements. Note that the index $%
k $ is odd. Just like $b_{1}$ the symmetric matrix $\overline{b}_{k1}$ has a
diagonal element at position ($2^{n-1}-1,2^{n-1}-1$) and hence the operator $%
\overline{b}_{k1}$ also contains the diagonal operator $D_{2^{n-1}-1}.$ The
diagonal operator $D_{2^{n-1}-1}$ is commutable with both the operators $%
b_{1}$ and $\overline{b}_{k1}$. The unitary operator $\overline{B}%
_{k1}(\theta )=\exp (-i\theta \overline{b}_{k1})$ can be decomposed
efficiently. Just like the subspace-selective multiple-quantum operator (27)
(see next section) the Hermitian operator $\overline{b}_{k1}$ can be
expressed as 
\begin{eqnarray*}
(\overline{b}_{k1}-D_{2^{n-1}-1}) &=&[\overline{g}%
_{k},(b_{1}-D_{2^{n-1}-1})]_{+} \\
&=&\overline{g}_{k}(b_{1}-D_{2^{n-1}-1})+(b_{1}-D_{2^{n-1}-1})\overline{g}%
_{k}
\end{eqnarray*}
where the diagonal operator $\overline{g}_{k}$ is given by 
\[
\overline{g}%
_{k}=Diag(0_{0},...,0_{(k-1)/2-1};1_{(k-1)/2},...,1_{(2^{n-1}-1)-1};0_{(2^{n-1}-1)},...,0_{N-1}). 
\]
As shown in section 3.1, the unitary diagonal operator $\overline{G}%
_{k}(\theta )=\exp (-i\theta \overline{g}_{k})$ can be decomposed
efficiently into a sequence of the $l-$qubit selective rotation operations $%
C_{l,t_{m}}(\theta )$ $(l=1,2,...,$ $n)$ with number less than $2n$. Since
both the two operators $(\overline{b}_{k1}-D_{2^{n-1}-1})$ and $%
(b_{1}-D_{2^{n-1}-1})$ do not contain any diagonal operator components and
both the two operators $\overline{g}_{k}(b_{1}-D_{2^{n-1}-1})$ and $%
(b_{1}-D_{2^{n-1}-1})\overline{g}_{k}$ have not any overlapping nonzero
matrix element it follows from the unitary transformation (38) and the
decomposition formula (41) in next section that the unitary operator $%
\overline{B}_{k1}(\theta )=\exp (-i\theta \overline{b}_{k1})$ can be
efficiently decomposed as 
\begin{eqnarray}
\overline{B}_{k1}(\theta ) &=&C_{2^{n-1}-1}(\theta )\exp [-i\theta (%
\overline{b}_{k1}-D_{2^{n-1}-1})]  \nonumber \\
&=&C_{2^{n-1}-1}(\theta )(\exp [-i\frac{1}{2}\theta (b_{1}-D_{2^{n-1}-1})/L]%
\overline{G}_{k}(\pi )  \nonumber \\
&&\times \exp [i\frac{1}{2}\theta (b_{1}-D_{2^{n-1}-1})/L]\overline{G}%
_{k}(\pi )^{-1})^{L}+O(L^{-1})  \nonumber \\
&=&C_{2^{n-1}-1}(\theta )[\exp (-i\frac{1}{2}\theta b_{1}/L)\overline{G}%
_{k}(\pi )  \nonumber \\
&&\times \exp (i\frac{1}{2}\theta b_{1}/L)\overline{G}_{k}(\pi
)^{-1}]^{L}+O(L^{-1}).  \label{25}
\end{eqnarray}
For a modest number $L=\varepsilon ^{-1}$ this expansion of the unitary
operator $\overline{B}_{k1}(\theta )$ can fast converge. As shown in (b),
the unitary operator $\exp (-i\frac{1}{2}\theta b_{1}/L)$ is a sequence of $%
n $ basic unitary operations. Then the unitary operator $\overline{B}%
_{k1}(\theta )$ can be decomposed as a sequence of the basic unitary
operations with number less than $6\varepsilon ^{-1}n$.

It can be proved that the unitary operator $U_{k}$ in (24) with the index $%
k=2^{k_{l-1}}+2^{k_{l-2}}+...+2^{k_{1}}+1$\ ($n-1>k_{l-1}>k_{l-2}>...>k_{1}%
\geq 1$ and $1<l\leq n-1)$ can be written as 
\begin{equation}
U_{k}=\left\{ 
\begin{array}{l}
\exp (i\frac{\pi }{2}b_{j_{1}})\exp (i\frac{\pi }{2}b_{-j_{2}})\exp (i\frac{%
\pi }{2}b_{j_{3}}) \\ 
\times ...\times \exp (i\frac{\pi }{2}b_{-j_{l-1}}),\text{ if }l-1\text{ is
even,} \\ 
\exp (i\frac{\pi }{2}b_{j_{1}})\exp (i\frac{\pi }{2}b_{-j_{2}})\exp (i\frac{%
\pi }{2}b_{j_{3}}) \\ 
\times ...\times \exp (i\frac{\pi }{2}b_{j_{l-1}})\exp (i\frac{\pi }{2}%
b_{0}),\text{ if }l-1\text{ is odd,}
\end{array}
\right.  \label{26}
\end{equation}
where the indices $j_{1}=2^{k_{l-1}-1},$ $%
j_{2}=2^{k_{l-2}-1},...,j_{l-1}=2^{k_{1}-1}.$ Since the unitary operator $%
\exp (-i\theta b_{\pm j})$ with $j=2^{l}$ $(l=1,2,...,n-1)$ can be
decomposed as a sequence of $(n-l)$ basic unitary operations, as shown in
(c), then the unitary operator $\exp (i\frac{\pi }{2}b_{\pm j_{m}})$ with
index $j_{m}=2^{k_{l-m}-1}$ in (26) can be decomposed into a sequence of $%
(n-k_{l-m}+1)$ basic unitary operations. Therefore, number of the basic
unitary operations in the unitary operator $U_{k}$ (26) is $%
(n-k_{l-1}+1)+(n-k_{l-2}+1)+...+(n-k_{1}+1)$ if $l-1$ is even or $%
(n-k_{l-1}+1)+(n-k_{l-2}+1)+...+(n-k_{1}+1)+1$ if $l-1$ is odd. Note that $%
(n-k_{l-1}+1)+(n-k_{l-2}+1)+...+(n-k_{1}+1)+1<(l-1)n+1<n^{2}.$ The unitary
operator $U_{k}$ (26) then can be decomposed into a sequence of the basic
unitary operations with number less than $n^{2}.$

Therefore, the expansion (25) of the unitary operator $\exp (-i\theta 
\overline{b}_{k1})$ and the decomposition (26) of the unitary operator $%
U_{k} $ show that the unitary operator $B_{k}(\theta )$ (24) built up with a
general anti-diagonal operator $b_{k}$ can be expressed as a sequence of the
basic unitary operations with number less than $2n^{2}+6\varepsilon ^{-1}n$,
and quantum-circuit complexity for the unitary operator is $%
O(2n^{2}+6\varepsilon ^{-1}n)$.\newline
\newline
\newline
{\large 3.3 The subspace-selective multiple-quantum operators}

The Hermitian multiple-quantum operator $Q_{pm}$ in the subspace-selective
multiple-quantum unitary operator $U_{pm}(\theta )=\exp (-i\theta Q_{pm})$
(5) can be generated by the anti-commutator of the diagonal and
anti-diagonal operators: 
\begin{equation}
Q_{pm}=\stackrel{d(m)-1}{\stackunder{l=0}{\sum }}Q_{pml}=\frac{1}{2}[%
b_{k},g_{m}]_{+}=\frac{1}{2}(g_{m}b_{k}+b_{k}g_{m}),  \label{27}
\end{equation}
where the anti-diagonal operator $b_{k}$ needs to be chosen properly and the
diagonal operator $g_{m}$ $\in S_{zq}(m)\times S_{zq}(m)$ so that the $p-$%
quantum unitary operator $U_{pm}(\theta )$ built up with the Hermitian $p-$%
quantum operator $Q_{pm}$ is selectively applied on both the state subspace $%
S_{zq}(m)$ and another larger subspace $S_{zq}(m+p)$. It needs first to show
how to choose the anti-diagonal operator $b_{k}$ to generate the Hamiterian
operator $Q_{pm}.$ Here consider only the case $0\leq m<n/2$. For the case $%
n\geq m>n/2$ the multiple-quantum operator $Q_{pm}$ can be constructed with
the operator $b_{-k}$ in place of the operator $b_{k}$ and the final result
is similar. For convenience the anti-diagonal operator $b_{k}\equiv
(b_{k})_{N\times N}$ $(N=2^{n})$ is written as 
\[
b_{k}\equiv
Adiag(1_{[0,N-k-1]},1_{[1,N-k-2]},...,1_{[N-k-1,0]};0_{[N-k,N-1]},...,0_{[N-1,N-k]}). 
\]
Suppose that the operator $b_{k}$ is chosen properly so that the operator $%
g_{m}b_{k}$ is given by 
\begin{eqnarray*}
g_{m}b_{k}
&=&Diag(0_{0},...,0_{l_{m}-1};1_{l_{m}},...,1_{L_{m}};0_{L_{m}+1},...,0_{N-1})
\\
&&\times Adiag(1_{[0,N-k-1]},1_{[1,N-k-2]},...,1_{[N-k-1,0]}; \\
&&0_{[N-k,N-1]},...,0_{[N-1,N-k]}) \\
&=&Adiag(0_{[0,N-k-1]},...,0_{[l_{m}-1,N-k-1-l_{m}+1]}; \\
&&1_{[l_{m},N-k-1-l_{m}]},...,1_{[L_{m},N-k-1-L_{m}]}; \\
&&0_{[L_{m}+1,N-k-1-L_{m}-1]},...,0_{[N-k-1,0]};0_{[N-k,N-1]},...,0_{[N-1,N-k]}),
\end{eqnarray*}
and the operator is written as 
\begin{eqnarray*}
b_{k}g_{m}
&=&Adiag(1_{[0,N-k-1]},1_{[1,N-k-2]},...,1_{[N-k-1,0]};0_{[N-k,N-1]},...,0_{[N-1,N-k]})
\\
&&\times
Diag(0_{0},...,0_{l_{m}-1};1_{l_{m}},...,1_{L_{m}};0_{L_{m}+1},...,0_{N-1})
\\
&=&Adiag(0_{[0,N-k-1]},...,0_{[N-k-1-L_{m}-1,L_{m}-1]}; \\
&&1_{[N-k-1-L_{m},L_{m}]},...,1_{[N-k-1-l_{m},l_{m}]}; \\
&&0_{[N-k-1-l_{m}+1,l_{m}-1]},...,0_{[N-k-1,0]};0_{[N-k,N-1]},...,0_{[N-1,N-k]}).
\end{eqnarray*}
Obviously, the index $N-1-k$ must be greater than the index $L_{m}$. Both
the operators $g_{m}b_{k}$ and $b_{k}g_{m}$ are also anti-diagonal
operators. Their nonzero elements taking one are also along the same
anti-diagonal line of the matrix $b_{k}$. Each of the two operators has only 
$d(m)$ nonzero matrix elements, which number is exactly dimensional size of
the state subspace $S_{zq}(m)$. The operator $(g_{m}b_{k}+b_{k}g_{m})$ is
clearly a symmetric and Hermitian operator and can be written as 
\begin{eqnarray*}
&&(g_{m}b_{k}+b_{k}g_{m}) \\
&=&Adiag(0_{[0,N-k-1]},...,0_{[l_{m}-1,N-k-1-l_{m}+1]}; \\
&&1_{[l_{m},N-k-1-l_{m}]},0_{[l_{m}+1,N-k-1-l_{m}-1]},...,0_{[N-k-1-l_{m}-1,l_{m}+1]};
\\
&&1_{[N-k-1-l_{m},l_{m}]};0_{[N-k-1-l_{m}+1,l_{m}-1]},...,0_{[N-k-1,0]}; \\
&&0_{[N-k,N-1]},...,0_{[N-1,N-k]}) \\
&&+Adiag(0_{[0,N-k-1]},...,0_{[l_{m},N-k-1-l_{m}]}, \\
&&1_{[l_{m}+1,N-k-1-l_{m}-1]},0_{[l_{m}+2,N-k-1-l_{m}-2]},...,0_{[N-k-1-l_{m}-2,l_{m}+2]},
\\
&&1_{[N-k-1-l_{m}-1,l_{m}+1]},0_{[N-k-1-l_{m},l_{m}]},...,0_{[N-k-1,0]}; \\
&&0_{[N-k,N-1]},...,0_{[N-1,N-k]})+...... \\
&&+Adiag(0_{[0,N-k-1]},...,0_{[L_{m}-1,N-k-1-L_{m}+1]}, \\
&&1_{[L_{m},N-k-1-L_{m}]};0_{[L_{m}+1,N-k-1-L_{m}-1]},...,0_{[N-k-1-L_{m}-1,L_{m}+1]},
\\
&&1_{[N-k-1-L_{m},L_{m}]};0_{[N-k-1-L_{m}+1,L_{m}-1]},...,0_{[N-k-1,0]}; \\
&&0_{[N-k,N-1]},...,0_{[N-1,N-k]}).
\end{eqnarray*}
Therefore, the operator $(g_{m}b_{k}+b_{k}g_{m})$ can be expressed as a sum
of $d(m)$ commutative and symmetric anti-diagonal operators at most, each of
which has only two elements taking one along the same anti-diagonal line of
the matrix $b_{k}$. Such an anti-diagonal operator can be written in terms
of the usual computational basis $\{|\varphi _{k}\rangle \}:$%
\begin{eqnarray*}
&&Adiag(0_{[0,N-k-1]},...,0_{[l_{m}+l^{\prime }-1,N-k-1-l_{m}-l^{\prime
}+1]}, \\
&&1_{[l_{m}+l^{\prime },N-k-1-l_{m}-l^{\prime }]},0_{[l_{m}+l^{\prime
}+1,N-k-1-l_{m}-l^{\prime }-1]},...,0_{[N-k-1-l_{m}-l^{\prime
}-1,l_{m}+l^{\prime }+1]}, \\
&&1_{[N-k-1-l_{m}-l^{\prime },l_{m}+l^{\prime }]},0_{[N-k-1-l_{m}-l^{\prime
}+1,l_{m}+l^{\prime }-1]},...,0_{[N-k-1,0]}; \\
&&0_{[N-k,N-1]},...,0_{[N-1,N-k]}) \\
&=&|\varphi _{l_{m}+l^{\prime }}\rangle \langle \varphi
_{N-k-1-l_{m}-l^{\prime }}|+|\varphi _{N-k-1-l_{m}-l^{\prime }}\rangle
\langle \varphi _{l_{m}+l^{\prime }}|,\text{ }
\end{eqnarray*}
where the index $l^{\prime }=0,1,...,L_{m}-l_{m}$ $(L_{m}-l_{m}=d(m)-1)$ and
the base $|\varphi _{l_{m}+l^{\prime }}\rangle $ belongs to the state
subspace $S_{zq}(m)$. If both the operators $g_{m}b_{k}$ and $b_{k}g_{m}$
have common nonzero matrix elements then the operator $%
(g_{m}b_{k}+b_{k}g_{m})$ is a sum of commutative and Hermitian anti-diagonal
operators with number less than $d(m)$. If the basis state $|\varphi
_{N-k-1-l_{m}-l^{\prime }}\rangle $ belongs to another subspace $S(M_{m}+p)$
this anti-diagonal operator is a state-selective $p-$quantum operator that
applies only on both the two basis states $|\varphi _{l_{m}+l^{\prime
}}\rangle $ and $|\varphi _{N-k-1-l_{m}-l^{\prime }}\rangle $. Now it is
required that both the operators $g_{m}b_{k}$ and $b_{k}g_{m}$ have not any
common nonzero matrix elements so that the operator $(g_{m}b_{k}+b_{k}g_{m})$
is a sum of the commutative anti-diagonal operators with number $d(m)$
exactly. Then the index $N-1-k>2L_{m},$ and the operator $%
(g_{m}b_{k}+b_{k}g_{m})$ can be written in terms of the usual computational
basis 
\begin{eqnarray}
&&\frac{1}{2}(g_{m}b_{k}+b_{k}g_{m})  \nonumber \\
&=&\stackrel{d(m)-1}{\stackunder{l^{\prime }=0}{\sum }}\frac{1}{2}(|\varphi
_{l_{m}+l^{\prime }}\rangle \langle \varphi _{N-k-1-l_{m}-l^{\prime
}}|+|\varphi _{N-k-1-l_{m}-l^{\prime }}\rangle \langle \varphi
_{l_{m}+l^{\prime }}|).  \label{28}
\end{eqnarray}
Note that all the $d(m)$\ basis states $|\varphi _{l_{m}+l^{\prime }}\rangle 
$ ($l^{\prime }=0,1,...,d(m)-1$) in (28) belong to the subspace $S_{zq}(m)$.
If now all the $d(m)$ basis states $|\varphi _{N-k-1-l_{m}-l^{\prime
}}\rangle $ in (28) belong to the subspace $S_{zq}(m+p)$ whose dimensional
size is larger than that of $S_{zq}(m)$, the operator $\frac{1}{2}%
(g_{m}b_{k}+b_{k}g_{m})$ is a subspace-selective multiple-quantum operator
which applies only to both the two subspaces $S_{zq}(m)$ and $S_{zq}(m+p)$.
This can be seen more clearly by comparing the operator (28) with the
operator (27) and the multiple-quantum operators (3) and (5). Therefore, the
condition that the operator $\frac{1}{2}(g_{m}b_{k}+b_{k}g_{m})$ is the
subspace-selective $p-$quantum operator $Q_{pm}$ (5) selectively applied to
both the subspaces $S_{zq}(m)$ and $S_{zq}(m+p)$ is that in addition to $%
N-1-k>2L_{m}$ the index $k$ for the anti-diagonal operator $b_{k}$ must
satisfy 
\begin{equation}
l_{m+p}\leq N-k-1-l_{m}-l^{\prime }\leq L_{m+p},\text{ }l^{\prime
}=0,1,...,d(m)-1.  \label{29}
\end{equation}
Note that $L_{m}=l_{m}+d(m)-1$ and $L_{m}<l_{m+p}$ for $0\leq m+p\leq n/2$
and $p\geq 1.$ The condition $N-1-k>2L_{m}$ always holds if the condition $%
L_{m}<l_{m+p}\leq N-k-1-l_{m}-(d(m)-1)$ holds. Thus, the condition (29) is a
general condition to determine the proper index $k$ for the anti-diagonal
operator $b_{k}$ that is used to construct the subspace-selective $p-$%
quantum operator $Q_{pm}$ (27).

Consider the $n-$qubit spin system with an even qubit number $n.$ Then the
largest subspace for the system is $S_{zq}(n/2).$ Let $%
S_{zq}(m+p)=S_{zq}(n/2).$ Then the condition (29) is rewritten as 
\begin{equation}
l_{n/2}\leq N-k-1-l_{m}-l^{\prime }\leq L_{n/2},\text{ }l^{\prime
}=0,1,...,d(m)-1.  \label{30}
\end{equation}
The condition (30) shows that $l_{n/2}\leq N-k-1-l_{m}\leq L_{n/2}$ if $%
l^{\prime }=0$ and if $l^{\prime }=d(m)-1$ then $l_{n/2}\leq N-k-1-L_{m}\leq
L_{n/2},$ and since $N-k-1-l_{m}>N-k-1-(l_{m}+1)>...>N-k-1-L_{m}$ one has
for $l^{\prime }=0,1,...,d(m)-1,$ 
\[
l_{n/2}\leq N-k-1-L_{m}<...<N-1-k-l_{m}\leq L_{n/2}. 
\]
Therefore, the lower bound $(k_{m})_{\min }$ and upper bound $(k_{m})_{\max
} $ for the index $k\equiv k_{m}$ of the subspace $S_{zq}(m)$ for $%
m=0,1,...,n/2-1$ are given by 
\[
(k_{m})_{\min }=N-l_{m}-L_{n/2}-1\text{ and }(k_{m})_{\max
}=N-L_{m}-l_{n/2}-1, 
\]
and hence the index $k_{m}$ is bounded on by 
\begin{equation}
N-l_{m}-l_{n/2}-d(n/2)\leq k_{m}\leq N-l_{m}-l_{n/2}-d(m).  \label{31}
\end{equation}
The condition (31) shows that range of the index $k_{m}$ is equal to $\Delta
k_{m}=(k_{m})_{\max }-(k_{m})_{\min }=d(n/2)-d(m).$ For a different state
subspace $S_{zq}(m)$ the distance $\Delta k_{m}$ is different, and the
maximum and minimum $\Delta k_{m}$ are $\Delta k_{0}=d(n/2)-d(0)=\left( 
\begin{array}{l}
n \\ 
n/2
\end{array}
\right) -1$ and $\Delta k_{n/2-1}=d(n/2)-d(n/2-1)=\frac{2}{n+2}\left( 
\begin{array}{l}
n \\ 
n/2
\end{array}
\right) ,$ respectively, and moreover $\Delta k_{0}>\Delta k_{1}>...>\Delta
k_{n/2-1}.$ Note that $2l_{n/2}+d(n/2)=\sum_{m=0}^{n}d(m)=\sum_{m=0}^{n}%
\left( 
\begin{array}{l}
n \\ 
m
\end{array}
\right) =2^{n}=N.$ The condition (31) is reduced to the form 
\begin{equation}
l_{n/2}-l_{m}\leq k_{m}\leq l_{n/2}-l_{m}+d(n/2)-d(m)  \label{32}
\end{equation}
where $l_{m}=d(0)+d(1)+...+d(m-1)$ and $l_{0}=0,$ and $%
l_{n/2}>l_{n/2-1}>...>l_{1}>l_{0}.$ Now using the condition (32) one can
determine the index $k_{m}$ for the desired operator $b_{k_{m}}.$ Suppose
that the index $k_{m}=2^{n-1}.$ Since $N/2=l_{n/2}+d(n/2)/2>l_{n/2}$ the
first inequality in the condition (32) always holds: $%
k_{m}=N/2>l_{n/2}-l_{m}.$ The second inequality is reduced to the form 
\begin{equation}
d(n/2)/2-l_{m}-d(m)\geq 0.  \label{33}
\end{equation}
Therefore, the operator $b_{k_{m}}$ with $k_{m}=2^{n-1}$ can be used to
construct those $p-$quantum operators $Q_{pm}$ with index $m=0,1,...,m_{0}$
where the maximum index $m_{0}$ is determined from the inequality (33): 
\[
l_{m_{0}+1}=\stackrel{m_{0}}{\stackunder{l=0}{\sum }}d(l)\leq \frac{1}{2}%
d(n/2). 
\]
As shown in section 3.2, the anti-diagonal operator $b_{k_{m}}$ with index $%
k_{m}=2^{n-1}$ can be explicitly expressed as 
\[
b_{k_{m}}=(\frac{1}{2}E_{1}+I_{1z})\bigotimes 2^{n-1}I_{2x}I_{3x}...I_{nx} 
\]
and the corresponding unitary operator $B_{k_{m}}(\theta )$ therefore is a
basic unitary operation, 
\begin{equation}
B_{k_{m}}(\theta )=\exp [-i\theta (\frac{1}{2}E_{1}+I_{1z})\bigotimes
2^{n-1}I_{2x}I_{3x}...I_{nx}].  \label{34}
\end{equation}
This quantum unitary operator is used to build up a subspace-selective $p-$%
quantum unitary operator $U_{pm}(\theta )$ (5) that can transfer any unknown
state that is in one of the subspaces $S_{zq}(0)$, $S_{zq}(1)$,..., $%
S_{zq}(m_{0})$ into the largest subspace $S_{zq}(n/2)$.

Next consider the situation $m_{0}<m\leq n/2-1.$ Since $%
l_{n/2}-l_{m}=d(m)+d(m+1)+...+d(n/2-1)$ the condition (32) is rewritten as 
\begin{equation}
d(m)+d(m+1)+...+d(n/2-1)\leq k_{m}\leq d(m+1)+d(m+2)+...+d(n/2)\newline
.  \label{35}
\end{equation}
In particular, for $m=n/2-1$ the condition (35) is written as 
\begin{equation}
\left( 
\begin{array}{l}
n \\ 
n/2-1
\end{array}
\right) \leq k_{n/2-1}\leq \left( 
\begin{array}{l}
n \\ 
n/2
\end{array}
\right) .  \label{36}
\end{equation}
By the minimum distance $\Delta k_{n/2-1}=\frac{2}{n+2}\left( 
\begin{array}{l}
n \\ 
n/2
\end{array}
\right) $ one may obtain the index $k_{m}$ that satisfies (35) for each
subspace $S_{zq}(m)$ with $m_{0}<m\leq n/2-1$. The minimum distance $\Delta
k_{n/2-1}$ is approximated by using the Starling$^{\prime }s$ formula $%
n!\thickapprox \sqrt{2\pi n}(n/e)^{n}$ for a large $n,$%
\[
\Delta k_{n/2-1}=\frac{2}{n+2}\left( 
\begin{array}{l}
n \\ 
n/2
\end{array}
\right) \thickapprox \frac{2}{n+2}2^{n}\frac{\sqrt{2}}{\sqrt{\pi n}} 
\]
and 
\[
\log _{2}\Delta k_{n/2-1}\thickapprox n-\log _{2}(\frac{1}{2}n\sqrt{n}\sqrt{%
\frac{\pi }{2}}(1+\frac{2}{n})). 
\]
Denote $n_{0}=\lfloor \log _{2}(\frac{1}{2}n\sqrt{n}\sqrt{\frac{\pi }{2}}(1+%
\frac{2}{n}))\rfloor $ as integer part of $\log _{2}(\frac{1}{2}n\sqrt{n}%
\sqrt{\frac{\pi }{2}}(1+\frac{2}{n})).$ Then $2^{n-n_{0}-1}\leq \Delta
k_{n/2-1}\leq 2^{n-n_{0}}$ and $2^{n_{0}}\sim n^{3/2}.$ The minimum index $%
(k_{n/2-1})_{\min }$ that satisfies (36) is approximated by the Starling
formula 
\[
(k_{n/2-1})_{\min }=\left( 
\begin{array}{l}
n \\ 
n/2-1
\end{array}
\right) \thickapprox \sqrt{\frac{2}{\pi }}\frac{n}{n+2}\frac{1}{\sqrt{n}}%
2^{n} 
\]
and 
\[
\log _{2}\left( 
\begin{array}{l}
n \\ 
n/2-1
\end{array}
\right) \thickapprox n-\log _{2}(\sqrt{n}\sqrt{\frac{\pi }{2}}(1+\frac{2}{n}%
)). 
\]
Let $k_{0}=\lfloor \log _{2}(\sqrt{n}\sqrt{\frac{\pi }{2}}\frac{1}{1+2/n}%
)\rfloor .$ Then $2^{n-k_{0}-1}\leq (k_{n/2-1})_{\min }\leq 2^{n-k_{0}}$ and 
$2^{k_{0}}\sim n^{1/2}.$ Although taking $k_{n/2-1}=2^{n-k_{0}}$ the first
inequality in (36) can be satisfied, the second inequality in (36) could not
be satisfied. Now suppose that the index $k_{n/2-1}=2^{n-k_{0}-1}+\mu
2^{n-n_{0}-1},$ $\mu $ is an integer to be determined. Then there is an
integer $\mu $ such that the index $k_{n/2-1}$ satisfies the condition (36).
There is always an integer $\mu _{0}$ such that $2^{n-k_{0}-1}+\mu
_{0}2^{n-n_{0}-1}<(k_{n/2-1})_{\min }\leq $ $2^{n-k_{0}-1}+(\mu
_{0}+1)2^{n-n_{0}-1}.$ The integer $\mu _{0}$ equals $\lfloor
((k_{n/2-1})_{\min }-2^{n-k_{0}-1})/2^{n-n_{0}-1}\rfloor .$ Because $%
2^{n-k_{0}-1}+\mu _{0}2^{n-n_{0}-1}\leq (k_{n/2-1})_{\min }$ and $%
2^{n-n_{0}-1}\leq \Delta k_{n/2-1}$ there must be $2^{n-k_{0}-1}+(\mu
_{0}+1)2^{n-n_{0}-1}\leq (k_{n/2-1})_{\min }+\Delta k_{n/2-1}=d(n/2).$ By
taking $\mu =\mu _{0}+1$ the index $k_{n/2-1}=2^{n-k_{0}-1}+\mu
2^{n-n_{0}-1} $ satisfies the condition (36). Generally, suppose that there
is an integer $\mu _{0}$ such that $2^{n-k_{0}-1}+\mu
_{0}2^{n-n_{0}-1}<d(m)+d(m+1)+...+d(n/2-1)\leq $ $2^{n-k_{0}-1}+(\mu
_{0}+1)2^{n-n_{0}-1}.$ Because $2^{n-n_{0}-1}\leq \Delta k_{n/2-1}<\Delta
k_{m}$ there holds $2^{n-k_{0}-1}+(\mu
_{0}+1)2^{n-n_{0}-1}<d(m)+d(m+1)+...+d(n/2-1)+\Delta
k_{m}=d(m+1)+d(m+1)+...+d(n/2).$ Therefore, the index $k_{m}=$ $%
2^{n-k_{0}-1}+\mu _{m}2^{n-n_{0}-1}$ with $\mu _{m}=\mu _{0}+1$ satisfies
the condition (35). For $m_{0}<m\leq n/2-1$ the upper bound for the index $%
k_{m}$ is obtained from the condition (35) by taking $m=m_{0}+1:$%
\begin{eqnarray*}
k_{m} &\leq &d(m_{0}+2)+d(m_{0}+3)+...+d(n/2) \\
&=&l_{n/2}+d(n/2)-d(0)-d(1)-...-d(m_{0})-d(m_{0}+1) \\
&<&N/2.
\end{eqnarray*}
This is because $\stackrel{m_{0}+1}{\stackunder{l=0}{\sum }}d(l)>\frac{1}{2}%
d(n/2)$ and $\stackrel{m_{0}}{\stackunder{l=0}{\sum }}d(l)\leq \frac{1}{2}%
d(n/2),$ as shown in (33). Since $k_{m}<2^{n-1}$ the index $k_{m}$ can be
expanded as a binary number: 
\begin{eqnarray}
k_{m} &=&2^{n-k_{0}-1}+\mu _{m}2^{n-n_{0}-1}  \nonumber \\
&=&a_{n-2}2^{n-2}+a_{n-3}2^{n-3}+...+a_{n-n_{0}-1}2^{n-n_{0}-1}  \label{37}
\end{eqnarray}
where $a_{l}=0,1$ $(l=n-n_{0}-1,n-n_{0},...,n-2).$ The anti-diagonal
operator $b_{k_{m}}$ with the index (37) then can be expressed as, as can be
seen in (22), 
\[
(b_{k_{m}})_{2^{n}\times 2^{n}}=(b_{k_{m}^{\prime }})_{2^{n_{0}+1}\times
2^{n_{0}+1}}\bigotimes 2^{n-n_{0}-1}I_{n_{0}+2x}I_{n_{0}+3x}...I_{nx} 
\]
where $k_{m}^{\prime
}=a_{n-2}2^{n_{0}-1}+a_{n-3}2^{n_{0}-2}+...+a_{n-n_{0}-1}2^{0}$ and the
anti-diagonal operator $b_{k_{m}^{\prime }}$ has a dimensional size $%
2^{n_{0}+1}\times 2^{n_{0}+1}.$ Because $2^{n_{0}}\sim n^{3/2}$ the operator 
$(b_{k_{m}^{\prime }})_{2^{n_{0}+1}\times 2^{n_{0}+1}}$ can be expressed as
a sum of $\sim n^{3/2}$ commutative product operators at most using the
recursive relation (23) in section 3.2. Therefore, the operator $b_{k_{m}}$
is also a sum of $\sim n^{3/2}$ commutative product operators at most.

Furthermore, if using the general decomposition (24) for a general
anti-diagonal operator the unitary operator $\exp (-i\theta b_{k_{m}}$) with
the index $k_{m}$ (37) can be decomposed into a sequence of the basic
unitary operations with a complexity only $O(2(n_{0}+1)^{2}+6\varepsilon
^{-1}(n_{0}+1)),$ which is $\thicksim O(\log _{2}^{2}n)$.

For an $n-$qubit spin system with an odd qubit number the largest subspaces
are $S_{zq}(n/2-1/2)$ and $S_{zq}(n/2+1/2)$. Any state in one of the two
largest subspaces can be converted unitarily into another largest subspace
by a unitary transformation with $n$ single-spin unitary operations: $%
U=\prod_{k=1}^{n}\exp (-i\pi I_{kx}).$ This unitary transformation is also
available for the complete quantum-state transfer between any pair of
symmetric subspaces $S_{zq}(m)$ and $S_{zq}(n-m)$ in a general $n-$qubit
spin system. Thus, it needs only to consider those subspaces $S_{zq}(k_{m})$
with $0\leq k_{m}\leq (n-1)/2.$ Take $S_{zq}(m+p)=S_{zq}(n/2-1/2)$. Now the
minimum distance $\Delta k_{(n-1)/2-1}=d(n/2-1/2)-d((n-1)/2-1)=\frac{4}{n+3}%
\left( 
\begin{array}{l}
n \\ 
(n-1)/2
\end{array}
\right) $ and is approximated by 
\begin{eqnarray*}
\log _{2}\Delta k_{(n-1)/2-1} &\thickapprox &n-\log _{2}\{\sqrt{\frac{\pi }{2%
}}\frac{n+3}{4}\frac{(1+n)\sqrt{n-1}}{n} \\
&&\times \sqrt{(1+\frac{1}{n})^{n}}/\sqrt{[1+\frac{1}{n-1}]^{(n-1)}}\}.
\end{eqnarray*}
Now let $n_{0}=\lfloor \log _{2}\sqrt{\frac{\pi }{2}}\frac{n+3}{4}\frac{(1+n)%
\sqrt{n-1}}{n}\sqrt{(1+\frac{1}{n})^{n}}/\sqrt{[1+\frac{1}{n-1}]^{(n-1)}}%
\rfloor .$ Note that $(1+\frac{1}{n})^{n}\thickapprox e$ for a large integer 
$n.$ Then $2^{n_{0}}\thicksim n^{3/2}.$ Therefore, for the case that the
qubit number $n$ is odd the operator $b_{k_{m}}$ with $k_{m}=2^{n-1}$ is
still used to construct those $p-$quantum operators $Q_{pm}$ (27) with index 
$m=0,1,...,m_{0},$ but now the maximum index $m_{0}$ is determined from the
inequality: $\stackrel{m_{0}}{\stackunder{l=0}{\sum }}d(l)\leq \frac{1}{2}%
d((n-1)/2),$ and when $m_{0}<m<n/2-1/2$ the index $k_{m}$ is still given by
(37). The complexity of quantum circuit of the unitary operator $\exp
(-i\theta b_{k_{m}})$ is also $\thicksim O(\log _{2}^{2}n)$.

For a general subspace $S_{zq}(m+p)$ instead of the largest subspace in an $%
n-$qubit spin system the index $k$ of the operator $b_{k\newline
}$ used to construct the $p-$quantum operator $Q_{pm}$ (27) is generally
determined from the general condition (29), and the unitary operator $\exp
(-i\theta b_{k})$ is decomposed as a sequence of the basic unitary
operations by the general decomposition (24), and the complexity of quantum
circuit of the unitary operator $\exp (-i\theta b_{k})$ is $%
O(2n^{2}+6n\varepsilon ^{-1})$.

Once the unitary operator $\exp (-i\theta b_{k_{m}})$ with the anti-diagonal
operator $b_{k_{m}}$is expressed as an efficient sequence of the basic
unitary operations the subspace-selective $p-$quantum unitary operator $%
U_{pm}(\theta )$ (5) can be easily decomposed as an efficient sequence of
the basic unitary operations$.$ A general unitary transformation with a
sequence of selective rotation operations $\{C_{k}(\theta _{k})\}$ can help
simplify the decomposition. The unitary transformation has been given in
Ref.[6]: 
\[
U_{o}(\theta _{0},\theta _{1},...,\theta _{m-1})\rho _{I}(t)U_{o}(\theta
_{0},\theta _{1},...,\theta _{m-1})^{-1} 
\]
\[
=\rho _{I}(t)-[\rho _{I}(t),\stackrel{m-1}{\stackunder{k=0}{\sum }}(1-\cos
\theta _{k})D_{k}]_{+}+i[\rho _{I}(t),\stackrel{m-1}{\stackunder{k=0}{\sum }}%
D_{k}\sin \theta _{k}] 
\]
\[
+\stackrel{m-1}{\stackunder{k=0}{\sum }}\stackrel{m-1}{\stackunder{l=0}{\sum 
}}[(1-\cos \theta _{k})(1-\cos \theta _{l})+\sin \theta _{k}\sin \theta
_{l}]D_{k}\rho _{I}(t)D_{l} 
\]
\begin{equation}
+\stackrel{m-1}{\stackunder{l>k=0}{\sum }}i[\sin \theta _{k}(1-\cos \theta
_{l})-\sin \theta _{l}(1-\cos \theta _{k})](D_{k}\rho _{I}(t)D_{l}-D_{l}\rho
_{I}(t)D_{k})  \label{38}
\end{equation}
\newline
where the diagonal unitary operation $U_{o}(\theta _{0},\theta
_{1},...,\theta _{m^{\prime }-1})=\stackrel{m^{\prime }-1}{\stackunder{k=0}{%
\prod }}C_{k}(\theta _{k}).$ Now setting $U_{o}(\theta _{0},\theta
_{1},...,\theta _{m^{\prime }-1})=G_{m}(\pi )$ and $\rho _{I}(t)=b_{k}$ and
inserting them into the unitary transformation (38) one obtains 
\begin{eqnarray}
&&G_{m}(\pi )b_{k}G_{m}(\pi )^{-1}  \nonumber \\
&=&b_{k}-2[b_{k},g_{m}]_{+}+4\stackrel{d(m)-1}{\stackunder{k^{\prime }}{\sum 
}}\stackrel{d(m)-1}{\stackunder{l^{\prime }}{\sum }}D_{k^{\prime
}}b_{k}D_{l^{\prime }},\text{ }  \label{39}
\end{eqnarray}
where $D_{k^{\prime }},D_{l^{\prime }}\in S_{zq}(m)\times S_{zq}(m).$ Since
an anti-diagonal operator $b_{k}$ with an even index $k,$ e.g., $k_{m}$ does
not contain any diagonal operator component, that is, all the diagonal
elements of the matrix $b_{k}$ are equal to zero, there must be $%
D_{k^{\prime }}b_{k}D_{k^{\prime }}=(b_{k})_{k^{\prime }k^{\prime
}}D_{k^{\prime }}=0$ for any diagonal operator $D_{k^{\prime }}.$ On the
other hand, there also holds: $D_{k^{\prime }}b_{k_{m}}D_{l^{\prime }}=0$ $%
(k^{\prime }\neq l^{\prime }),$ $D_{k^{\prime }},D_{l^{\prime }}\in
S_{zq}(m)\times S_{zq}(m)$ $(0\leq m<n/2).$ The matrix element of $%
D_{k^{\prime }}b_{k}D_{l^{\prime }}$ is given by 
\[
(D_{k^{\prime }}b_{k}D_{l^{\prime }})_{ij}=\sum_{t}\sum_{s}\delta
_{ik^{\prime }}\delta _{tk^{\prime }}(b_{k})_{ts}\delta _{sl^{\prime
}}\delta _{jl^{\prime }}=\delta _{ik^{\prime }}(b_{k})_{k^{\prime }l^{\prime
}}\delta _{jl^{\prime }}. 
\]
The element is not zero only when $i=k^{\prime }$ and $j=l^{\prime }$ and
the indices $k^{\prime }$ and $l^{\prime }$ satisfy the anti-diagonal line
equation of the matrix $b_{k}:k^{\prime }=-l^{\prime }+N-1-k.$ Since $%
D_{k^{\prime }},D_{l^{\prime }}\in S_{zq}(m)\times S_{zq}(m)$ the indices $%
k^{\prime }$ and $l^{\prime }$ satisfy $l_{m}\leq k^{\prime },l^{\prime
}\leq L_{m}$ and $k^{\prime }+l^{\prime }\leq 2L_{m}.$ However, it follows
from the anti-diagonal line equation that $k^{\prime }+l^{\prime
}=N-1-k>2L_{m}$ because the condition (29) shows that the index $%
N-1-k>2L_{m}.$ Therefore, the operator $D_{k^{\prime }}b_{k_{m}}D_{l^{\prime
}}$ is zero for those anti-diagonal operators $b_{k_{m}}$ used to build up
the $p-$quantum operator $Q_{pm}$. Then the unitary transformation (39) can
be further reduced to the form 
\begin{equation}
G_{m}(\pi )b_{k_{m}}G_{m}(\pi )^{-1}=b_{k_{m}}-2[b_{k_{m}},g_{m}]_{+}.
\label{40}
\end{equation}
With the help of the unitary transformation (40) and the Trotter-Suzuki
formula [13, 14] the subspace-selective $p-$quantum unitary operator $%
U_{pm}(\theta )=\exp (-i\theta Q_{pm})$ can be decomposed as 
\begin{eqnarray}
U_{pm}(\theta ) &=&\exp (-i\theta \frac{1}{2}[g_{m},b_{k_{m}}]_{+}) 
\nonumber \\
&=&\exp [-i\frac{1}{4}\theta (b_{k_{m}}-G_{m}(\pi )b_{k_{m}}G_{m}(\pi
)^{-1})]  \nonumber \\
&=&[\exp (-i\frac{1}{4}\theta b_{k_{m}}/L)G_{m}(\pi )  \nonumber \\
&&\times \exp (i\frac{1}{4}\theta b_{k_{m}}/L)G_{m}(\pi
)^{-1}]^{L}+O(L^{-1}).  \label{41}
\end{eqnarray}
Note that norm for the operators $b_{k_{m}}$ and $g_{m}$ and their
commutator $[b_{k_{m}},g_{m}]$ equals one, that is, $\left\|
b_{k_{m}}\right\| =1$, $\left\| g_{m}\right\| =1,$ and $\left\|
[b_{k_{m}},g_{m}]\right\| =1.$ Then for a modest number $L=\varepsilon ^{-1}$
the expansion (41) for the $p-$quantum unitary operator $U_{pm}(\theta )$
can converge quickly. The computational complexity for the quantum circuit
of the $p-$quantum unitary operator $U_{pm}(\theta )$ is therefore dependent
on that of the unitary operations $B_{k_{m}}(\theta )$ and $G_{m}(\theta ).$
It is shown in section 3.1 that the unitary operation $G_{m}(\theta )$ can
be decomposed into a sequence of $2n$ basic unitary operations at most. For
the situation that an unknown state in a subspace is transferred into the
largest subspace of the Hilbert space of the $n-$qubit spin system the
complexity of quantum circuit of the unitary operator $B_{k_{m}}(\theta )$
is $\thicksim O(\log _{2}^{2}n),$ and for a general case that an unknown
state is transferred from a subspace into a larger subspace the complexity
is $O(2n^{2}+6n\varepsilon ^{-1}).$ Therefore, it follows from (41) that the
subspace-selective $p-$quantum unitary operator $U_{pm}(\theta )$ can be
expressed as a sequence of the basic unitary operations with complexity $%
O(2(2n^{2}+6n\varepsilon ^{-1})\varepsilon ^{-1}+4n\varepsilon ^{-1})$.%
\newline
\newline
{\Large 4. Discussion}

It has been shown that any unknown quantum state can be efficiently
transferred from a state subspace into a larger state subspace of the
Hilbert space of an $n-$qubit spin system by a subspace-selective
multiple-quantum unitary transformation, but the Grover quantum search
algorithm [4, 15, 16] shows indirectly that the inverse process usually is
hard one. This multiple-quantum transition process is similar to evolution
process from nonequilibrium state to equilibrium state in a closed quantum
system, although the former is a unitary process and the latter a
non-unitary and irreversible process [17]. This result might be helpful for
understanding nonequilibrium processes such as protein folding process in
nature where energy effect usually is not dominating from the point view of
quantum dynamics. By the efficient subspace-selective multiple-quantum
unitary transformation that can transfer efficiently any state from a small
subspace into the largest subspace of the Hilbert space the search space of
the quantum search problem can be reduced from the whole Hilbert space to
its largest subspace. With the help of the results in the paper and the
auxiliary oracle unitary operation it can be shown that the quantum search
algorithm [2, 6] based on quantum dynamics is at least as powerful as those
quantum search algorithms including the Grover quantum search algorithm [4,
16] and adiabatic quantum search algorithm [18] in a pure quantum-state
system because the former algorithm needs only to find which subspace the
marked state is in among the $(n+1)$ subspaces of the Hilbert space. The
diagonal and the anti-diagonal unitary operators will have an extensive
application in constructing efficient quantum circuits for permutation
operations of a symmetry group and the unitary operations of a cyclic group.%
\newline
\newline
{\Large References}\newline
1. R.R.Ernst, G.Bodenhausen, and A.Wokaun, \textit{Principles of Nuclear
Magnetic Resonance in One and Two Dimensions}, (Oxford University Press,
Oxford, 1987)\newline
2. X.Miao, Universal construction for the unsorted quantum search
algorithms, http://arXiv.org/abs/quant-ph/0101126 (2001)\newline
3. X.Miao, A prime factorization based on quantum dynamics in a spin
ensemble (I), http://arXiv.org/abs/quant-ph/0302143 (2003)\newline
4. L.K.Grover, Quantum mechanics helps in searching for a needle in a
haystack, Phys.Rev.Lett. 79, 325 (1997) \newline
5. X.Miao, Universal construction of unitary transformation of quantum
computation with one- and two-body interactions, \newline
http://arXiv.gov/abs/quant-ph/0003068 (2000) \newline
6. X.Miao, Solving the quantum search problem in polynomial time on an NMR
quantum computer, http://arXiv.org/abs/quant-ph/0206102 (2002)\newline
7. E.U.Condon and G.H.Shortley, \textit{The theory of atomic spectra},
Chapter 3, Cambridge University Press, 1935.\newline
8. M.E.Rose, \textit{Elementary theory of angular momentum}, Wiley, New
York, 1957.\newline
9. A.R.Edmonds, \textit{Angular momentum in quantum mechanics}, 2nd edn,
Princeton University Press, Princeton, 1974.\newline
10. L.I.Schiff, \textit{Quantum mechanics}, 3rd edn, McGraw-Hill, Inc. New
York, 1968\newline
11. X.Miao, Multiple-quantum operator algebra spaces and description for the
unitary time evolution of multilevel spin systems, Molec.Phys. 98, 625
(2000) \newline
12. C.H.Bennett, Logical reversibility of computation, IBM J.Res.Dev. 17,
525-532 (1973)\newline
13. H.F.Trotter, On the product of semigroups of operators,

Proc.Am.Math.Soc. 10, 545 (1959)\newline
14. M.Suzuki, Decomposition formulas of exponential operators and Lie
exponentials with some applications to quantum mechanics and statistical
physics, J.Math.Phys. 26, 601 (1985) \newline
15. C.H.Bennett, E.Bernstein, G.Brassard, and U.Vazirani, Strengths and
weeknesses of quantum computing, SIAM Journal on Computing, 26, 1510 (1997)%
\newline
16. G.Brassard, P.Hoyer, M.Mosca, and A.Tapp, Quantum amplitude
amplification and estimation, http://arXiv.org/abs/quant-ph/0005055 (2000)%
\newline
17. E.M.Lifshitz and L.P.Pitaevskii, \textit{Statistical Physics (1)}, 3rd
edn, translated by J.B.Sykes and M.J.Kearsley, Pergamon Press, New York,
1980, \newline
18. (a) E.Farhi and S.Gutmann, Analog analogue of digital quantum
computation, Phys.Rev. A, 57, 2403 (1998);

(b) E.Farhi, J.Goldstone, S.Gutman, and M.Sipser, Quantum computation by
adiabatic evolution, http://arXiv.org/abs/quant-ph/0001106 (2000)

\end{document}